# Sex Differences in Severity and Mortality Among Patients With COVID-19: Evidence from Pooled Literature Analysis and Insights from Integrated Bioinformatic Analysis

Xiyi Wei[1,2,#], Yu-Tian Xiao[1,#], Jian Wang[1,3,6,#], Rui Chen[1,#], Wei Zhang[1,#], Yue Yang[1], Daojun Lv[4], Chao Qin[2], Di Gu[4], Bo Zhang[5], Weidong Chen[5], Jianquan Hou[3], Ninghong Song[2], Guohua Zeng[4], Shancheng Ren[1,*]

## Author Affiliations:

1. Department of Urology, Shanghai Changhai Hospital, Second Military Medical University, Shanghai 200433, China.
2. Department of Urology, The First Affiliated Hospital of Nanjing Medical University, Nanjing, Jiangsu 210009, China.
3. Department of Urology, The First Affiliated Hospital of Soochow University, 188 Shizi Road, Suzhou, Jiangsu 215006, China.
4. Department of Urology, The First Affiliated Hospital of Guangzhou Medical University, Shanghai, China.
5. NovelBio Bio-Pharm Technology Co.,Ltd., Shanghai, China.
6. Department of Urology, The Affiliated Hospital of Jiangnan University, Wuxi, 214000, China.

\# These authors contributed equally.

\* Correspondence:

Shancheng Ren, MD, PhD.

Department of Urology, Shanghai Changhai Hospital, 168 Changhai Road, Yangpu District, Shanghai 200433, Shanghai, China.

Email: renshancheng@gmail.com

## Abstract

**Importance:**

There are reports that a sex disparity exists in outcomes in patients with COVID-19, with men having more severe cases and higher mortality than women.

**Objective:**

To conduct a meta-analysis of current studies that examined sex differences in severity and mortality in patients with COVID-19, and identify potential mechanisms underpinning these differences.

**Methods:**

We performed a systematic review to collate data from observational studies examining associations of sex differences with clinical outcomes of COVID-19. PubMed, Web of Science and four preprint servers were searched for relevant studies. Data were extracted and analyzed using meta-analysis where possible, with summary data presented otherwise. Publicly available bulk RNA sequencing (RNA-seq), single-cell RNA sequencing (scRNA-seq), and chromatin immunoprecipitation sequencing (ChIP-seq) data were analyzed to explore the potential mechanisms underlying the observed association.

**Results:**

39 studies met inclusion criteria, representing 77932 patients, of which 41510 (53.3%) were males. Men were at a markedly increased risk of developing severe cases compared with women (OR = 1.63; 95% Cl = 1.28-2.06). Furthermore, the pooled odds ratio (OR) of mortality for male group compared with the female group indicated significant higher mortality rate for male (OR = 1.71; 95% CI = 1.51-1.93). Subgroup analyses suggested that in patients with average age > 50 yr, the male had significantly higher severity rate than female (OR = 1.94; 95% Cl = 1.16-3.26). However, in patients with average age < 50 yr, the male group only exhibited a marginally increased severity rate compared with the female group (OR = 1.45; 95% Cl = 1.07-1.96). Data from scRNA-seq suggest that men have a higher amount of ACE2-expressing pulmonary alveolar type II cells than women. Sex-based immunological differences exist, with cytokines associated with cytokine release syndrome preferentially expressed in men, and those associated with defense and recovery from viral infection preferentially expressed in women. ScRNA-seq data of the prostate and testis revealed that these two organs might be potential targets for SARS-CoV-2 infection in the male population. The expression of androgen receptor (AR) is positively correlated with ACE2, and there is evidence that AR may directly regulate the expression of ACE2.

**Conclusions:**

This meta-analysis detected an increased severity and mortality rate in the male populations with COVID-19, which might be attributable to the sex-based differences in cellular compositions and immunological microenvironments of the lung. The host cell receptor ACE2 is likely regulated by AR signaling pathway, which is identified as a potential target for prevention and treatment of SARS-Cov-2 infections in men.

# Keywords

# Introduction

Coronavirus disease 2019 (COVID-19), ascribed to the severe acute respiratory syndrome coronavirus 2 (SARS-CoV-2), is an emerging outbreak globally since December 2019. The management of COVID-19 has become a current emergency public health event, which has attracted increasing attention. Early recognition, testing and isolation are the effective response to curb transmissions.[1,2] Health security capacities improving, global collaboration in relation to COVID-19 and current travel restrictions are necessary for global epidemic control.[3,4] Around 6.4-24.9% of patients infected with SARS-CoV-2 may progress to severe and even fatal acute respiratory distress syndrome (ARDS), 5-26.1% of patients need treatment in the intensive care unit (ICU), and the fatality rate of patients can reach 1.4-7.2%.[5-8] Older age, higher Sequential Organ Failure Assessment (SOFA) score, chronic comorbidities were the risk factors associated with the patients of the 2019-nCoV infection[9,10]. However, whether the sex difference is related to the risk factors for infection, severity, and mortality of COVID-19 is still lacking a comprehensive analysis based on the integration of new studies.[11]

Although some cases report support for sex differences to be associated with the onset and prognosis of COVID-19 pneumonia,[12] some studies do not support this observation. The Chinese Center for Disease Control and Prevention reports on 44672 confirmed cases showing that sex differences may be a risk factor for mortality, but there is no sex differences in morbidity.[7] Even more confirmed cases of female than male in South Korea are reported by the Korean Society of Infectious Diseases.[13] Therefore, a meta-analysis was performed to reveal the correlation between sex differences and the prevalence, severity and mortality of COVID-19 pneumonia. ACE2 as a receptor of SARS-CoV and spike protein can be primed by TMPRSS2 are exploited to entry into target cells, which play an vital role in coronavirus pneumonia infection.[14] We also performed an integrated bioinformatic analysis, leveraging data from bulk microarray, bulk RNA sequencing (RNA-seq), single-cell tRNA sequencing (scRNA-seq), and chromatin immunoprecipitation sequencing (ChIP-seq) to provide valuable clues of possible underlying mechanism.

# Materials and Methods

## Systematic review and Meta-analysis

### Search Strategy and Selection Criteria

Two of the investigators (W.X. and W.J.) independently retrieved relevant literature from the databases including PubMed, Web of Science, bioRxiv and MedRxiv preprint servers from inception to March 19, 2020. We searched studies with no language restrictions to analyze the association between sex difference with prevalence, severity and mortality of the patients with the 2019-nCoV infection. Keywords and relative variant search terms were utilized as follows: SARS-CoV-2 and COVID-19. These terms were combined with "AND" or "OR". We first reviewed the titles and abstracts of the retrieved citations (excluding conference abstracts and critical articles), and then evaluated the eligibility for inclusion of the full text that was deemed pertinent.

We searched the literature on cases of the 2019-nCoV infection reported in different countries since December 2019. We excluded studies that may overlap based on the relative information reported in the studies, such as the time of patient data collection, hospital units and departments. We uniformity defined ICU patients, ARDS patients, severe patients and critical patients as severe group in our study.

Inclusion criteria should comply with the following:1) observational studies that involved patients with the 2019-nCoV infection had a detailed description of the sex ratio, number of severe or dead patients; 2) studies with the most reported cases if data overlap with another document. Exclusion Criteria: 1) studies that did not provide available full text; 2) studies not reported separately for gender or no desired data; 3) data overlapping.

## Data Extraction and Quality Assessment

Two investigators extracted independently the data from eligible studies. Any discrepancy was resolved by consensus. The following information was abstracted: the last name of the first author; date of manuscript acceptance or publication; date of patient collection; total patients in the study; number of male participants, number of female participants; number of severe cases; number of severe male cases; number of severe female cases; number of cases with comorbidities; number of deaths; number of male deaths; number of female deaths and smoking history.

We evaluated potential sources of bias in included studies using the appraisal tool to assess the quality of cross-sectional studies (AXIS) (Supplementary Table S1) .[15]

## Pooled Analysis

The data of individual studies were summarized. We attempted to perform analysis of sex difference predisposition of incidence, severity, and mortality in patients COVID-19. Subgroup analyses of sex differences predisposition of severity were also carried out according to the average age of patients in included studies. Meta-analysis was undertaken with Stata v12.0 (Stata Corporation, College Station, TX, USA) software, the results of which were presented in forest plots. Heterogeneity among studies were estimated using fixed-effects and random-effects models, which was reported using the Cochrane's Q-test [16] and the inconsistency index value ($I^2$) [17]. Funnel plots and Egger test were used to assess for publication bias and small study effects. All hypothesis tests were two-tailed. P < 0.05 was considered statistically significant.

For any outcome measurement that is otherwise not meta-analyzable, we presented an albatross plot,[18] which is a novel graphical tool for presenting results from studies, allowing an approximation of underlying effect sizes and the potential identification of heterogeneity sources across studies. The albatross plots were constructed using Stata v14.0 (Stata Corporation, College Station, TX, USA).

# Bioinformatic Analyses

## Single-cell RNA Sequencing Analysis Pipeline

Single-cell RNA Sequencing (scRNA-seq) data analysis was performed by NovelBio Bio-Pharm Technology Co.,Ltd. with NovelBrain Cloud Analysis Platform. We applied fastp [19] with default parameter filtering the adaptor sequence and removed the low quality reads to achieve the clean data based on the GSO datasets (Normal Prostate: GSE117403 and Normal Lung tissue: GSE122960). We downloaded the aggregated matrix. Cells contained over 200 expressed genes and mitochondria UMI rate below 20% passed the cell quality filtering and mitochondria genes were removed in the expression table.

Seurat package (version: 2.3.4, https://satijalab.org/seurat/) was used for cell normalization and regression based on the expression table according to the UMI counts of each sample and percent of mitochondria rate to obtain the scaled data. PCA was constructed based on the scaled data with top2000 high variable genes and top 10 principals of 20 PCs were used for tSNE construction and UMAP construction. Utilizing graph-based cluster method, we acquired the unsupervised cell cluster result based the PCA top 10 principal and we calculated the marker genes by FindAllMarkers function with wilcox rank sum test algorithm under following criteria:1. lnFC > 0.25; 2. P value<0.05; 3. min.pct>0.1.

The corresponding websites for the datasets, if available, were used for downloading figures for visualization of scRNA-seq data of normal human tissues of the prostate (https://strandlab.net/) and the testis (https://humantestisatlas.shinyapps.io/humantestisatlas1/).

## Cell Communication Analysis.

To enable a systematic analysis of cell–cell communication molecules, we applied cell communication analysis based on the CellPhoneDB,[20,21] a public repository of ligands, receptors and their interactions. Membrane, secreted and peripheral proteins of the cluster of different time point was annotated. Significant mean and Cell Communication significance (p value<0.05) was calculated based on the interaction and the normalized cell matrix achieved by Seurat Normalization.

## Differential Gene Expression Analysis

To identify differentially expressed genes among samples, the function FindMarkers with wilcox rank sum test algorithm was used under following criteria:1. lnFC > 0.25; 2. P value<0.05; 3. min.pct>0.1. Volcano plot was constructed by the R package "ggplot2".

## Co-regulation Network Analysis

To discover the gene co-regulation network, find gene modules function of monocle3 [22] was used with the default parameters.

## GO analysis

Gene ontology (GO) analysis was performed to elucidate the biological functions or processes of the differentially expressed gene in the experiment.[23] GO annotations were downloaded from NCBI (http://www.ncbi.nlm.nih.gov/), UniProt (http://www.uniprot.org/) and the Gene Ontology website (http://www.geneontology.org/). Fisher's exact test was utilized to identify the significant GO categories and q-values were calculated for false discovery rate control.

## Pathway Analysis

KEGG database was used to figure out the significant pathways enriched. Fisher's exact test was utilized to select the significant pathway, and the threshold of significance was defined by FDR cutoff values.[24]

## Bulk Microarray/RNA-seq Data Analysis

Microarray data available from public repositories were accessed from GSE5901 and GSE 56188 using the R package "GEOquery". Gene expression analysis and visualization were performed using custom R codes. Correlation between genes were analyzed using the publicly available TCGA datasets at the TIMER 2.0 website (http://timer.cistrome.org/).[25]

### ChIP-seq analysis

ChIP-seq data from the ENCODE project[26] were analyzed using the WashU EpiGenome Browser.[27] Human tissue samples or standard cell lines with both AR ChIP-seq data and histone mark ChIP-seq data available were selected.

# Results

## Study Selection and Characteristics

We recorded the selection process and completed a PRISMA flow diagram (Figure 1). From a total of 1561 records identified and screened by abstract and title, 108 articles were selected for full-text assessment. Among these, 69 were excluded due to lack of gender distinction or overlapping data. Thirty-nine studies, published between February 7, 2020 and March 17, 2020, were finally filtered for qualitative analysis and quantitative meta-analysis, thirty-four from China and five from other countries (Table 1), [5-8,10,13,28-60] including a total of 77932 patients with an approximate average age of 61.04. Most studies were retrospective study, and ten were descriptive case series. The comprehensive characteristics of patients subsumed are displayed in Table 1, including gender ratio, severe cases, smoking history, death cases and comorbidity information. Three variables were analyzed for the meta-analysis in our systematic review.

## Meta-analysis results

The comprehensive results of our meta-analysis are presented in Figure 2, Table 2, Table3 and Table 4. Thirty-nine studies involving 77932 patients were assessed in the comparison of the sex differences in patient composition. Part of the reported cases are concentrated in Wuhan, China, but from different hospitals. Overall, male accounted for 53.3% of all patients and female accounted for 46.7% (41510, 36408 respectively). Among the patients included, the number of males was higher than that of females (39 studies; 77932 patients; Odds = 1.12), especially in the Chinese population (34 studies; 50488 patients; Odds = 1.13). Because the sex-specific numbers of the general population in the investigated areas are typically not provided, a meta-analysis for sex-specific incidence or morbidity is not feasible. We therefore provided an albatross plot for visualization of the comparison of the male proportion in each study with 0.5 (Figure 2A). It can be observed from the albatross plot that studies with large sample sizes and low p values tend to report a male proportion over 0.5. Twenty-one studies (3905 patients) reported the severe cases according to the setting standards mentioned above. Regarding the severe case rate (Figure 2B), the male group exhibited a prominently increased severity rate compared with the female group (21 studies; 3905 patients; OR = 1.63; 95% Cl = 1.28-2.06, P=0.000). Subgroup meta-analysis suggested that in patients with average age > 50 yr, the male had significantly higher severity rate than female (OR = 1.94; 95% Cl = 1.16-3.26, P=0.000). However, in patients with average age < 50 yr, the male group only exhibited a marginally increased severity rate compared with the female group (OR = 1.45; 95% Cl = 1.07-1.96, P=0.006, Figure 2C). A total of 8 studies included information of smoking history with available information about percentage of severe cases.[8,10,39,43,44,55,58,60] With the percentage of positive smoking history ranged from 5.6% to 23.3%. A total of only 252 cases with smoking history were reported against the study outcome. We consider it would be inappropriate to pool the results for stratified analysis based on the limited cases. In the 39 studies mentioned above, a total of 8 studies (50936 patients) provided mortality information. The pooled estimate of OR of mortality for male group compared with the female group indicated significant survival risk for male (8 studies; 49869 patients; OR = 1.71; 95% CI = 1.51-1.93, P=0.000) (Figure 2D).

## Heterogeneity, Publication bias and Sensitivity analysis

Regarding the severity and mortality, moderate heterogeneity was observed ($I^2$=40.9%, $I^2$=0.0% respectively). However, heterogeneity increased when analysis of patients' sex composition was conducted (overall $I^2$=95.8%, China subgroup $I^2$=57.2%). Supplementary Figure S1A and S1B indicate that no significant publication bias for the gender composition and severe rate analysis was observed, which was confirmed by Eggers test (Table S2) (p=0.777, P = 0.055, respectively). In addition, there was no published bias for mortality analysis (Supplementary Figure S1C) (Egger's test: P = 0.376). The influence of each study on the combined results was detected by sensitivity analysis. Table S3 and Table S4 show that no individual study significantly affected the combined OR.

## Identification of Sex-Based Differential Gene Expression

To explore the molecular aspects of the potential sex-based physiological and immunological differences, we first analyzed a published human lung tissue scRNA-seq datasets. After data processing and quality control procedures, transcriptomic profiles for 43358 cells from human lung samples were acquired (Figure 3A).

We identified 19 clusters using graph-based clustering method. Based on marker gene expression, a total of 10 cell types were identified, including pulmonary alveolar type I (AT1) cells, AT2 cells, ciliated & goblet cells, endothelial cells, Fibroblasts, Macrophages, monocytes, plasma cells, and T cells. Comparison of the abundance of different cell types in male and female lung tissues was presented in Figure 3B. Most cell fractions do not differ greatly between tissues of male and female origins, except for plasma cells which are preferably enriched in female samples. Given the comparable percentages of total AT2 cells, men have a significantly higher percentage of ACE2-expressing AT2 cells. Feature plot also revealed a higher level of ACE2 expression in male lung tissue, especially in AT2 cells (Figure 3C). We performed a co-expression network analysis in the cells from male samples and discovered that, among others, androgen receptor (AR) exhibited a co-expression pattern with ACE2

(Figure 3D).

To further elucidate the role of sex-based impact on ACE2 expression and downstream effects, differential expression analysis was performed between ACE-expressing cells of male and female origins (Figure 4A). We discovered that the down-regulated genes in the male populations were enriched in pathways related to viral infections and immune response (Figure 4B). Intercellular communication analysis using CellPhoneDB revealed an active state of ACE-expressing AT2 cells, with ACE2 functioning both as a receptor and a ligand (Supplementary Figure S2). As cytokines are essential regulators of infection and immune response, we focused on the differential sex-based differential expression patterns on cytokines (Figure 4C). We discovered that IL6ST, a receptor of IL6, is expressed both in male and female population, with an average expression higher in male populations. Pro-inflammatory cytokines and chemokines, including CCL14, CCL23, IL7, IL16, and IL18, are also preferentially expressed in men, underlying the higher susceptibility of men developing cytokine release syndrome. TNFSF13B, which has been shown to be associated with the progression of chronic obstructive pulmonary disease (COPD) and pneumonia, is highly expressed in the monocyte-macrophage lineages of men. In contrast, cytokines with reported protective effects against viral infections, including CCL2, CCL3, and CCL4, are highly expressed in women.

In addition, we looked at scRNA-seq data of tissues from normal, male-specific organs including prostate and testis (Supplementary Figure S3-4). Two independent scRNA-seq datasets of the healthy human prostate tissues demonstrated that ACE2 is expressed in most prostate epithelial cell clusters, pericytes, and fibroblasts. scRNA-seq datasets of the testis revealed that ACE2 is preferentially expressed in Leydig and Sertoli cells. These data have provided evidence that prostate and testis could be potential targets for SARS-CoV-2 infection in men.

# Androgen receptor positively regulates ACE2 and TMPRSS2

Previous reports have highlighted the role of ACE2 and TMPRSS2 as dependencies for SARS-CoV-2 cell entry.[14] The role of AR acting as a transcription factor to activate the expression of TMPRSS2 has already been studied extensively in the context of various malignancies. We hypothesized that AR could also directly regulate ACE2 expression. At the single-cell level, we found a positive correlation between TMPRSS2 and ACE2 expression (Figure 5A, Pearson correlation coefficient r = 0.713). To verify this expression pattern across the spectrum of human tissues, we tested the correlation of expression level of AR and ACE2/TMPRSS2 in the TCGA PANCAN dataset (Figure 5B). We found that expression of ACE2 and TMPRSS2 is positively correlated with AR, irrespective of tissue types. We then utilized publicly available microarray datasets to investigate the impact of chemical or surgical castration, which lowers the level of androgens, on the expression level of these two genes. Unsurprisingly, both *in vitro* and *in vivo* studies (Figure 5C-D) demonstrated that castration leads to the decline of ACE2 and TMPRSS2, which could be reversed through androgen supplementation. To understand the mechanism underpinning the positive correlation between AR and ACE2, we analyzed publicly available ChIP-seq datasets and found that, around 4000 bp upstream of ACE2 transcription start site (TSS), there is overlap between AR ChIP-seq peaks and H3K4me1/H3K27ac ChIP-seq peaks (Supplementary Figure S5-6), indicating that AR likely binds to the enhancer regions and promotes the expression of ACE2.

# Discussion

In recent months, SARS-CoV-2 quickly becomes a serious public health issue worldwide. [59-62] COVID-19 is a species of coronavirus family, which is homologous with the SARS virus and MERS virus, causing diseases ranging from common cold to severe pneumonia. [63,64]However, the knowledge of the COVID-19 remains poorly understood. Altogether, the disease could be divided into four clinical types, including mild, moderate, severe and critical pneumonia.[7] Recently, the epidemiology and clinical features of patients with COVID-19 have been widely reported.[60,61,65,66] However, few studies have focused on the sex differences in the prevalence, severity and death of COVID-19.

This study is the first meta-analysis to appraise the role of gender in the incidence rate, morbidity and mortality of SARS-CoV-2 infection. Our pooled results demonstrated that gender played a prone role in COVID-19 infection. Concretely, male patients exhibited higher morbidity, incidence of severe disease and mortality compared with female patients. Of the 77992 patients with COVID-19 included in our study, 53.3% were men, indicating that men were more likely to be infected with sars-cov-2 than women. Unanimously, in the analysis of 1755 SARS cases by Karlberg et al.[67] the mortality rate of men was significantly higher than that of women (21.9% vs 13.2%, $P < 0.0001$). Leung et al,[68] showed that in SARS patients, men were more likely to experience adverse events. It is worth noting that there is no clear evidence of sex differences in the prognosis of influenza.[69-71] The sex differences of morbidity and prognosis is believed as a characteristic of coronavirus infection. In addition, we discovered that male patients with age > 50 yr were associated with a greater risk of progressing to severe cases or ICU cases. Interestingly, the epidemiological data from the 2002-2003 SARS epidemic and recent MERS also indicated that there might be sex and age-dependent differences in disease outcomes. According to the epidemiology analysis of SARS in the 2003 Hong Kong epidemic, both increasing age and male sex were associated with a greater risk for death.[68] Channappanavar et al. infected mice with SARS-CoV and demonstrated that male mice were more susceptible to SARS-CoV infection compared to age matched females. Furthermore, the degree of sex-bias to SARS-CoV infection in middle-aged mice was more pronounced compared to young mice.[72] Together, these data suggested that older male patients with COVID-19 should get more attention and prepared for ICU treatment when they were still mild cases.

The specific mechanism of sex differences is not clear. Some studies have shown that different outcomes between males and females may pertain to the possibility that estrogen protects women from worse clinical outcomes during SARS-CoV infection.[67,73] In a risk factor analysis of patients with COVID-19, smoking history was identified as a risk factor for disease progression.[43] It was illustrated that the expression of ACE2 is significantly higher in the lungs of smokers, which might be the reason for higher percentage of sever cases in smokers. Notably, the smoking rate of men is typically much higher than that of women.[12] This seems to explain the results theoretically, however, there was only a small proportion of cases with smoking history and thus the influence of smoking on sexual differences seems to be weak. Another possible explanation is the different levels of ACE2 expression. Current results show that SARS-CoV-2 enter into cells through ACE2, and the receptor binding domain of SARS-CoV-2 S and SARS-CoV possess homologous affinity to ACE2.[74] Research has shown that Compared with SARS-CoV, the receptor binding domain of SARS-CoV-2 has a higher binding force to ACE2 than that of SARS-CoV-2.[75] Anterior studies have reported that the high expression of ACE2 receptor in idiosyncratic organs of SARS patients pertains to the corresponding specific organ failure.[76,77] It should be noted that the ACE2 gene is pitched on the X chromosome. Studies have shown that men had higher levels of circulating ACE2 than women and patients with diabetes or cardiovascular disease.[78] Therefore, due to the high expression of ACE2, male patients may be more likely to develop severe symptoms and die of SARS-CoV-2. In addition to that, spike protein promotes the attachment and entry of coronavirus into target cells by binding to its cellular receptor.[79] Recent studies have confirmed that in COVID-19, SARS-CoV-2 also relies on the priming of spike protein by TMPRSS2 to enter target cells. Inhibitors of TMPRSS2, as a clinical option, can block virus entry.[14] However, the mechanisms need further elucidation.

In our study, single-cell transcriptomic profiling of human lung tissue samples revealed that ACE2 expression is higher in male lung tissue than in female, and that AR is co-expressed with ACE2/TMPRSS2. It is worth mentioning that androgen expression is significantly higher in men than women. This is consistent with our findings that COVID-19 has a higher incidence severity and mortality in men. We speculate that AR may be one of the important factors causing gender differences in COVID-19. Our further integrated bioinformatic analysis revealed that in LNCaP prostate cancer cell line, the expression of ACE2 and TMPRSS2 decreased significantly after treatment with the AR inhibitor bicalutamide. Similarly, in the prostate of castrated mice, the expression of ACE2 and TMPRSS2 decreased significantly, and their expression was up-regulated after supplementation with androgen. These evidences suggest that anti-AR may be a new strategy for treating males with SARS-CoV-2 infection. However, further research is needed to confirm this speculation.

A great quantity of patients with severe COVID-19 infection have experienced cytokine release syndromes, or "cytokine storms".[60] Previous cases of infectious diseases such as SARS, Middle East respiratory syndrome and Ebola virus infections have also demonstrated that cytokine storms can trigger the immune system to attack the body violently, which is a momentous cause of acute respiratory distress syndrome and multiple organ failure.[80-82] A previous study demonstrated that there was a positive correlation between the severity of pneumonia and the cytokine storm and inflammatory response caused by intravascular virus.[83,84] IL-6 is the main proinflammatory factor causing cytokine storm, which can significantly damage organ function, as well as IL-7, IL-16 and IL-18.[45,85,86] Our study observed that the expression levels of Il-6ST, IL-7, IL-16 and IL-18 in male were significantly higher than that in female. These findings and observations may help to explain why the risk of severe events in male is strikingly higher than in female.

Our data revealed that CCL14, CCL21, and CCL23 (important pro-inflammatory factors) have increased expression in the lungs of male patients with COVID-19, which may promote COVID-19-related cytokine storms. In flu disease, type I interferon-mediated production of the CCL2 recruits inflammatory dendritic cells (IDCs) to the tracheal epithelium, which is conducive to virus control.[87] Previous studies have shown that CCL3, CCL4 can inhibit HIV-1 from entering CD4 T cells and play a protective role.[88] CXCL16, mainly located in the respiratory epithelium, is a ligand for CXCR6 playing an important role in homeostasis of Resident memory T cells in the respiratory tract.[89] In our study, the expressions of CCL2, CCL3, CCL4, and CXCL16 were lower in male patients than female patients, which may aggravate the condition of male COVID-19 patients to a certain extent. BAFF (gene name Tnfsf13b) is closely related to B-cell activation and adaptive humoral immune response. BAFF is elevated in lung tissue of COPD patients and can cause lung inflammation and injury. [90] Therefore, our data show that BAFF expression is increased in male COVID-19 patients, which may worsen the condition.

Admittedly, there remain certain shortages in our systematic review. First of all, only thirty-nine studies were incorporated which focused on the morbidity, severity and mortality of COVID-19 infection. These studies are mostly from China. Up to March 28, 2020, more than 600,000 cases have been reported worldwide. Most of aforementioned cases have not been covered and published in the available studies. With an intensified body of case studies being reported, more large population and superior quality studies could be embedded to consolidate our findings. Secondly, our systematic review was consisted of 39 individual studies with various contexts, which may induce insufficient statistical availability and dispersive results to a lesser extent. Third, in rare cases, there is no way to determine which reports are repeated. there may be some overlap in patients among different studies, especially most of the included Chinese studies were summarized in the national report.[7] However, in the sensitivity analysis, the results were still consistent after omitting the Chinese national report in the analysis (Figure 4A). As for the heterogeneity of gender composition analysis, it may be attributed to a different baseline ratio of male and female in each region.

There may be sex differences in the susceptibility and diseases progression of patients with COVID-19. The clinical symptoms of male are more serious than those of women, and the outcome of hospitalization is less favorable. From the results of integrated bioinformatic analyses, it can be postulated that androgen deprivation therapy, frequently adopted in the treatment of prostate cancer, could potentially exert a protective and therapeutic effect against the coronavirus. Nevertheless, the findings of the study should be interpreted with caution. Large population-based studies are needed to verify these findings.

## Acknowledgments

**Funding:**

This research is partially funded by National Natural Science Foundation (81872105 to S.R.) and National Major R&D Program (2017YFC0908002 to S.R.)

**Conflicts of Interest:**

Z.B. and C.W. are co-founders of NovelBio Bio-Pharm Technology Co.,Ltd. The other authors have no conflicts of interest to disclose.

**Ethical Statement:**

The authors are accountable for all aspects of the work in ensuring that questions related to the accuracy or integrity of any part of the work are appropriately investigated and resolved.

# Figure Legends

**Figure 1. Flow diagram of the study selection process.**

**Figure 2. Visualization of pooled analyses of morbidity, severity, age subgroup severity and mortality.** (A) An albatross plot of the morbidity. (B) A forest plot of severity. (C) A forest plot of severity with age subgroup analysis. (D) A forest plot of mortality.

**Figure 3. Single-cell transcriptomic profiling of human lung tissue samples.** (A) t-SNE view of combined adult human lung tissue samples from 2 male and 6 female donors. Color coded by re-evaluated clusters. (B) Comparison of percentages of different cell types between lung tissue samples of male and female origins. (C) Feature plot showing the expression level of ACE2. (D) ACE2 co-expression network in lung tissues of male origin.

**Figure 4. Sex-based differences in gene expression, pathway enrichment and cell-cell communications.** (A) Volcano plot showing differentially expressed genes between male and female lung tissue samples. (B) Visualization of enriched pathways. (C) Bubble plot visualization of differentially expressed cytokines in ACE2-expressing cell populations.

**Figure 5. Co-expression analysis of AR, ACE2 and TMPRSS2.** (A) Single-cell RNAseq data of adult lung tissues showing a positive correlation between ACE2 and TMPRSS2 expression pattern. (B) Heatmap showing the correlation between AR and ACE2/TMPRSS2 in multiple cancer types across the TCGA PANCAN cohort. (C)(D) Changes in expression after chemical (C) or surgical (D) castration. r, Pearson correlation coefficient. N3, Prostate from sham-treated mouse. C3, Prostate, 3 days after castration. C14, Prostate, 14 days after castration. C14+T3, Prostate, 3 days after testosterone treatment of C14.

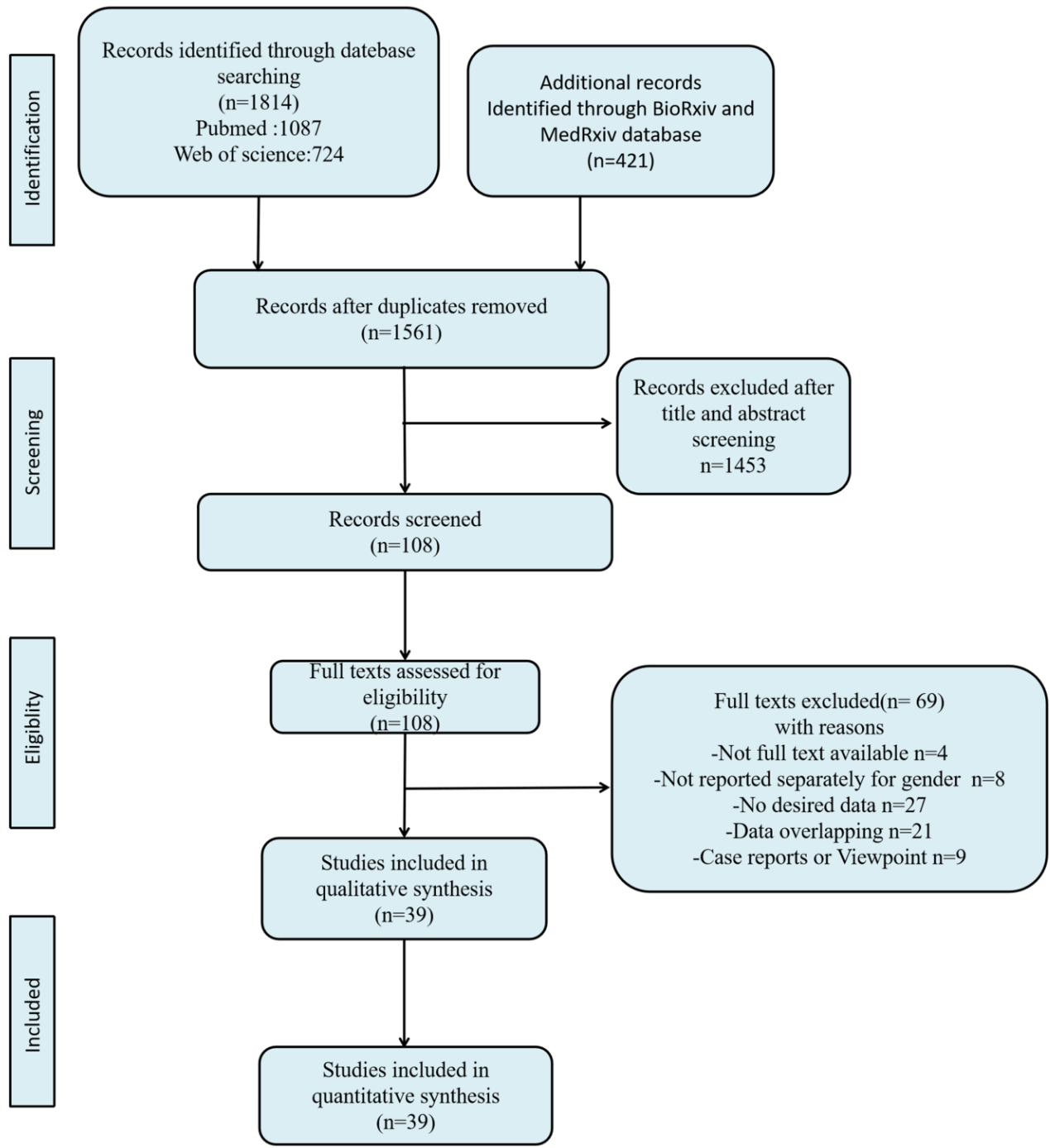

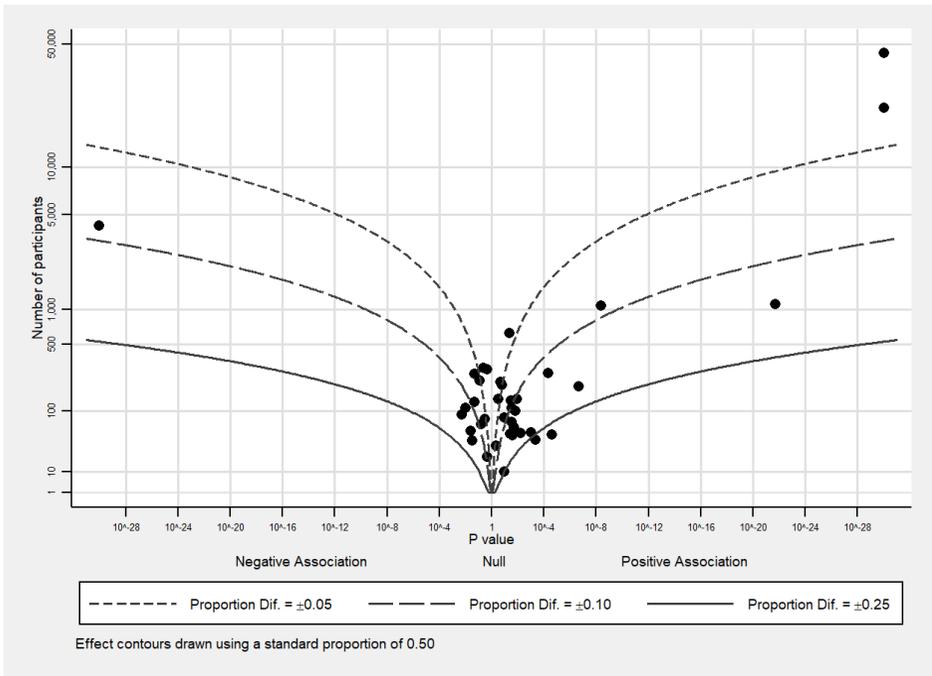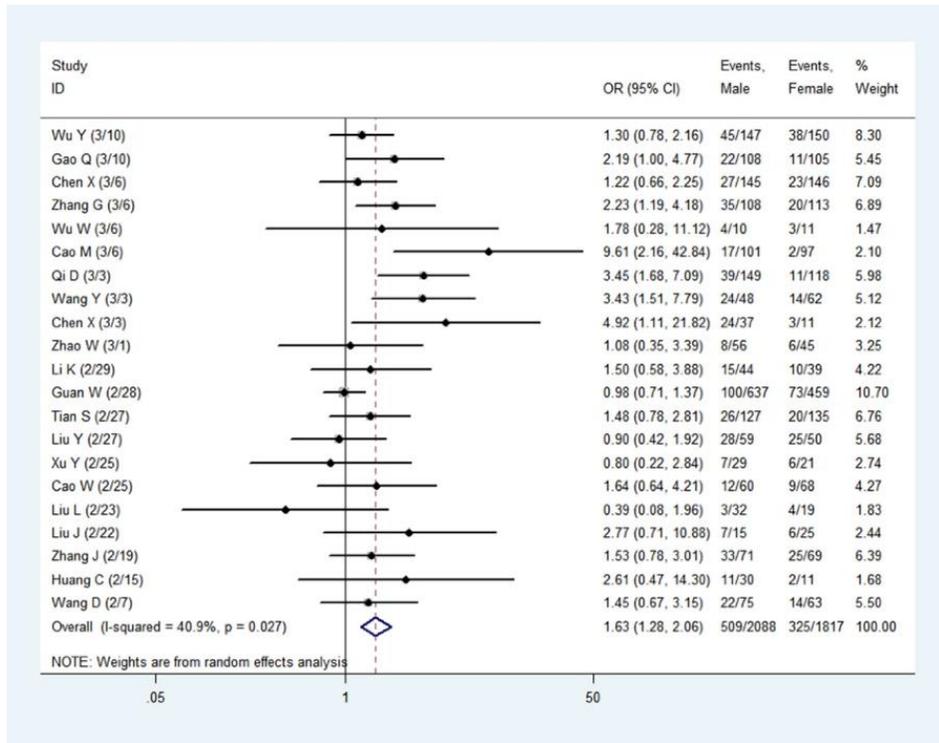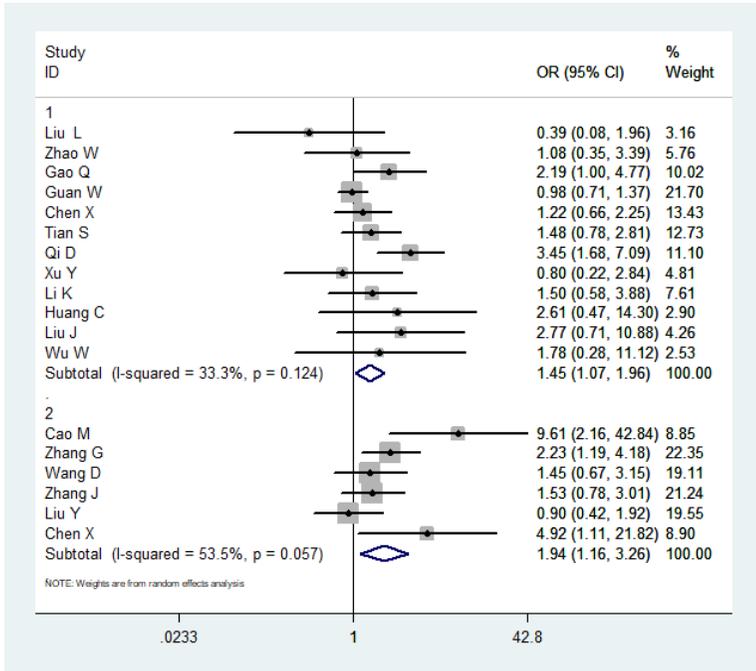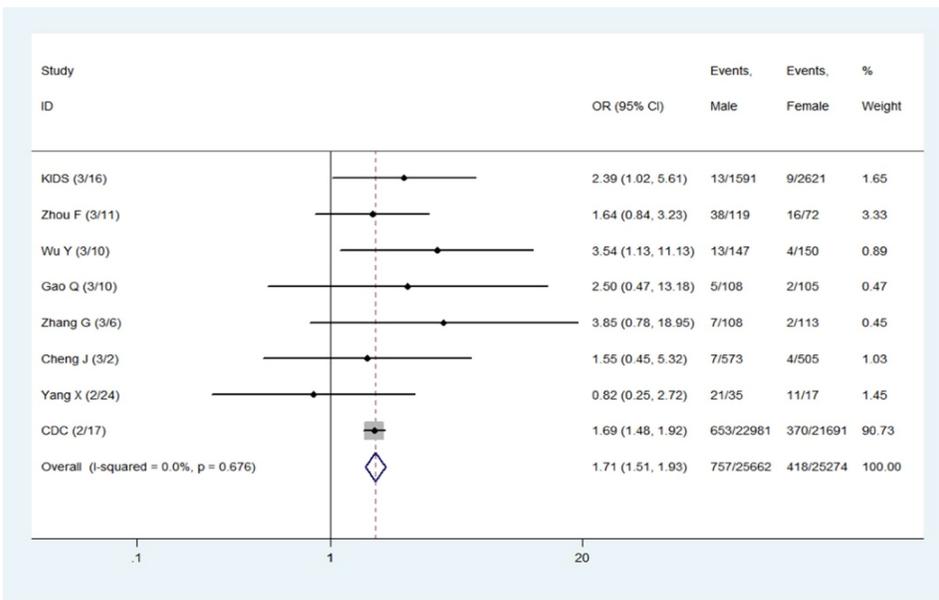

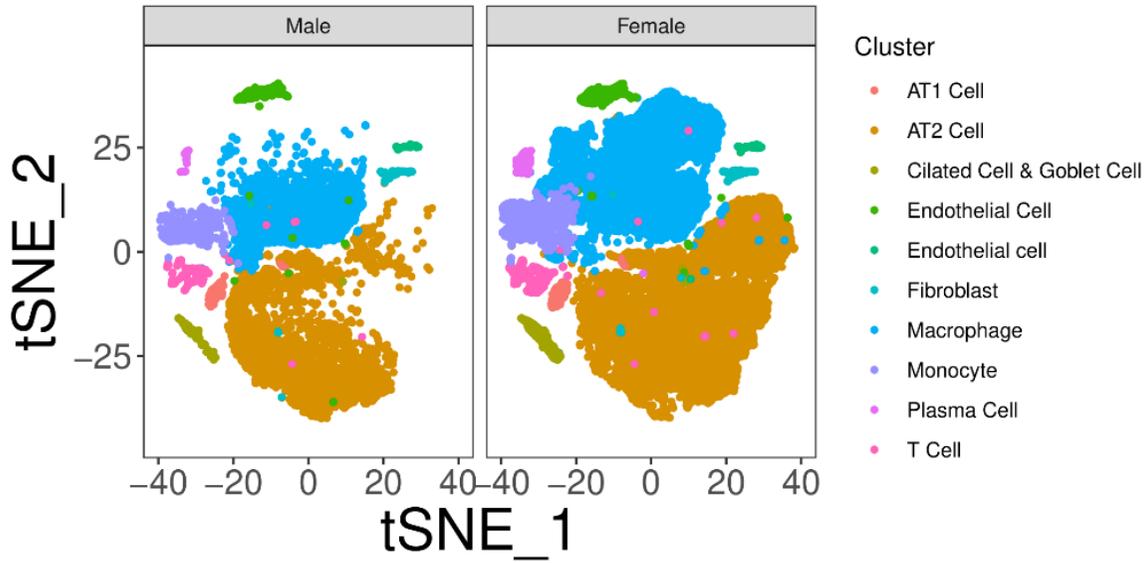
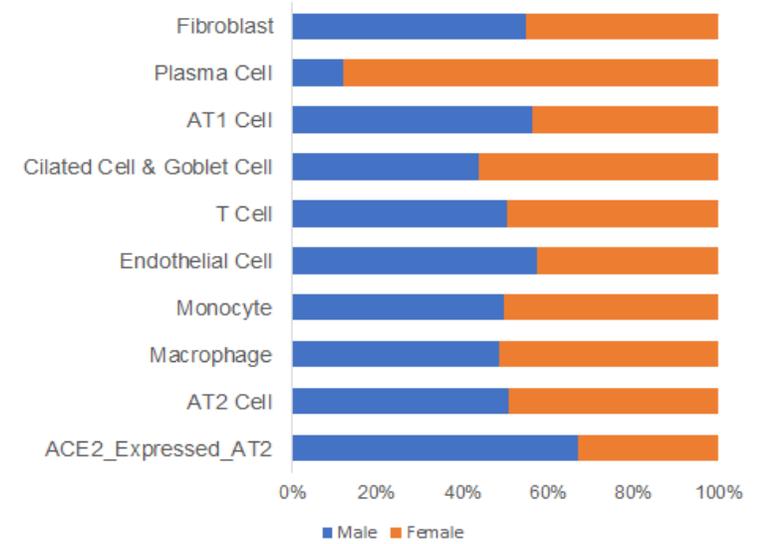
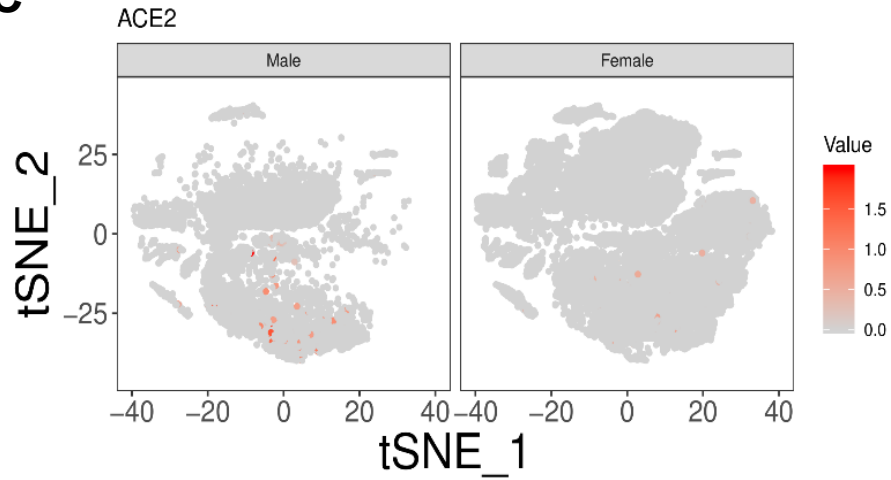
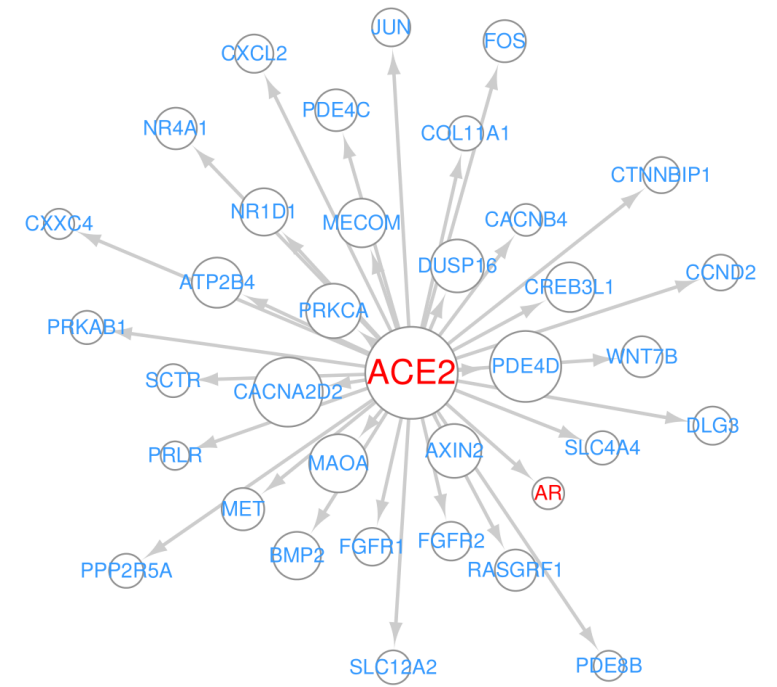

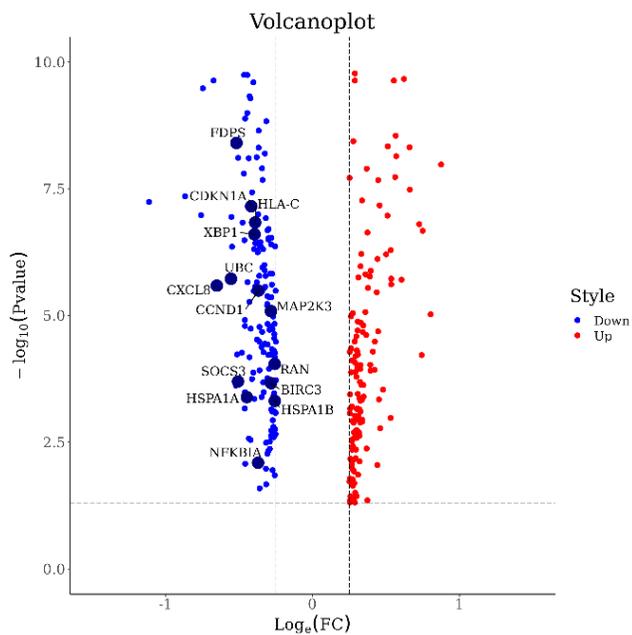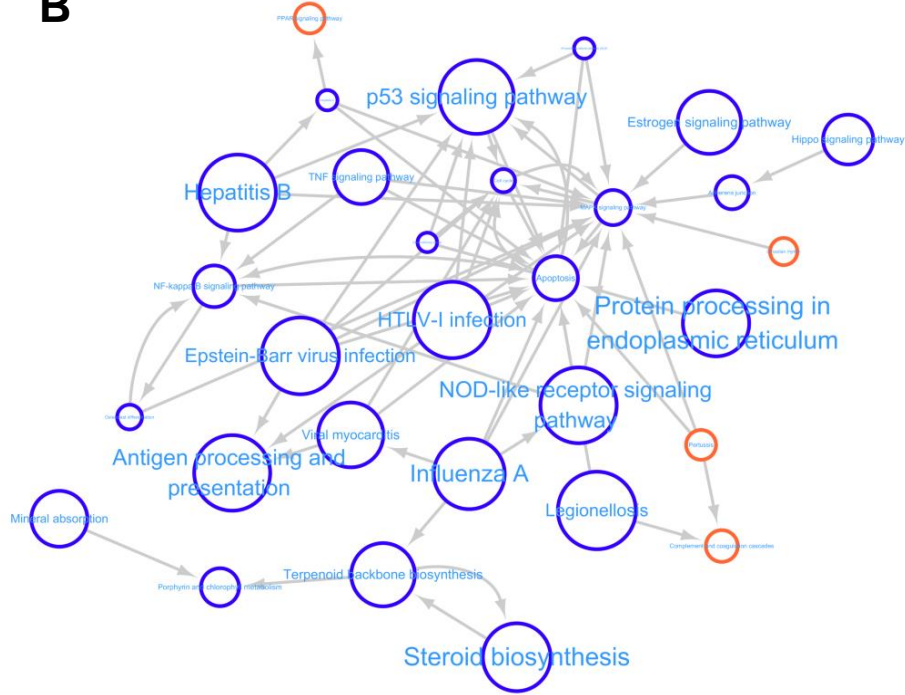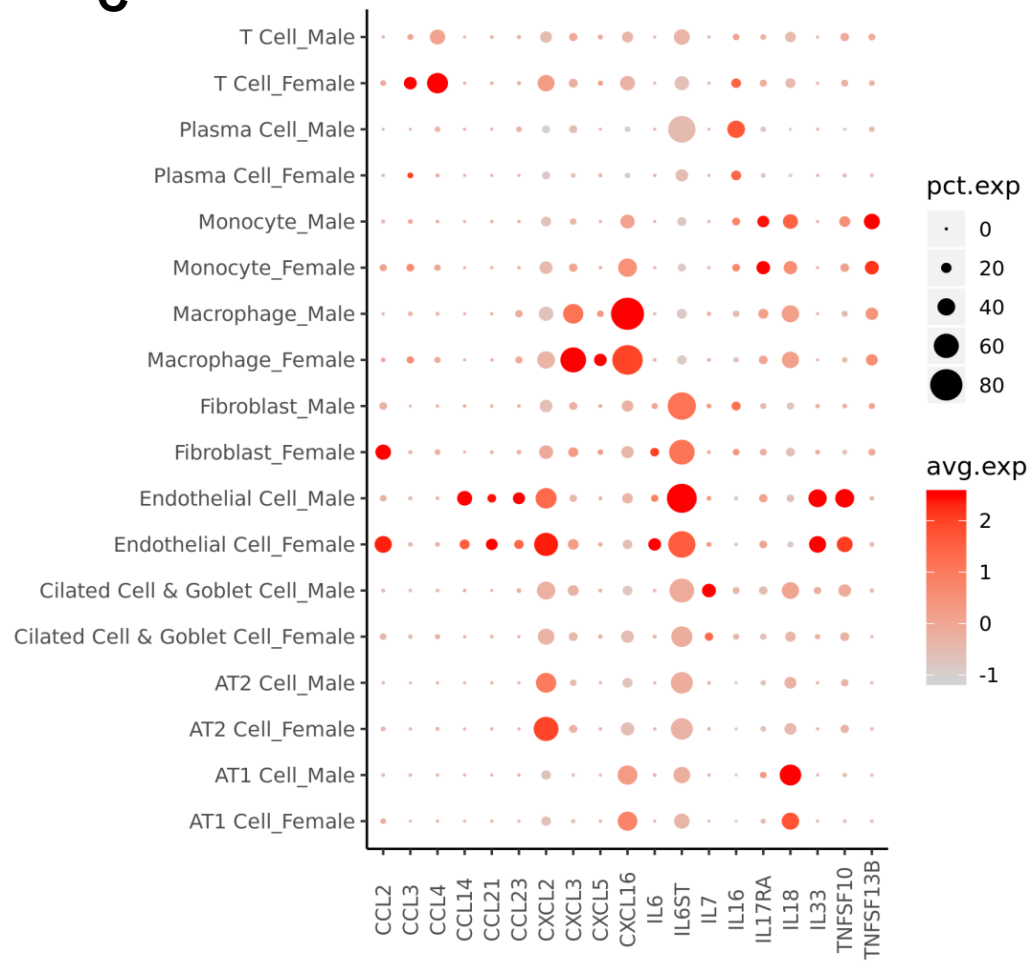

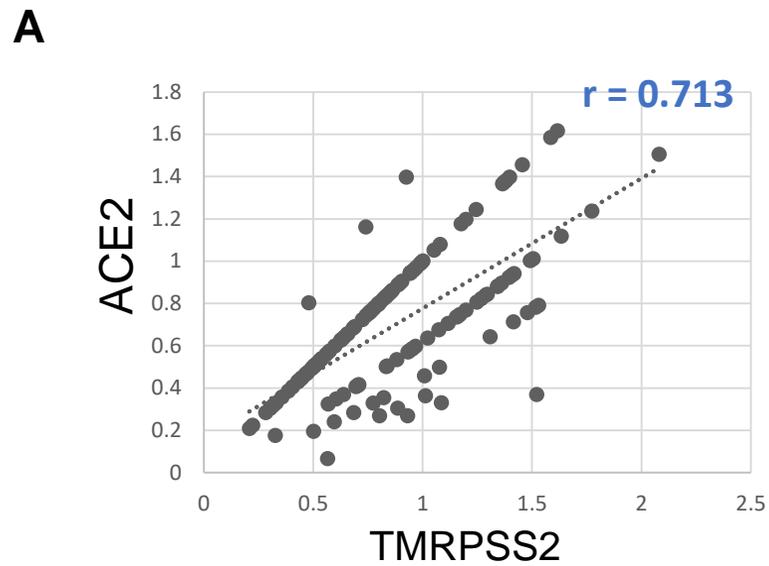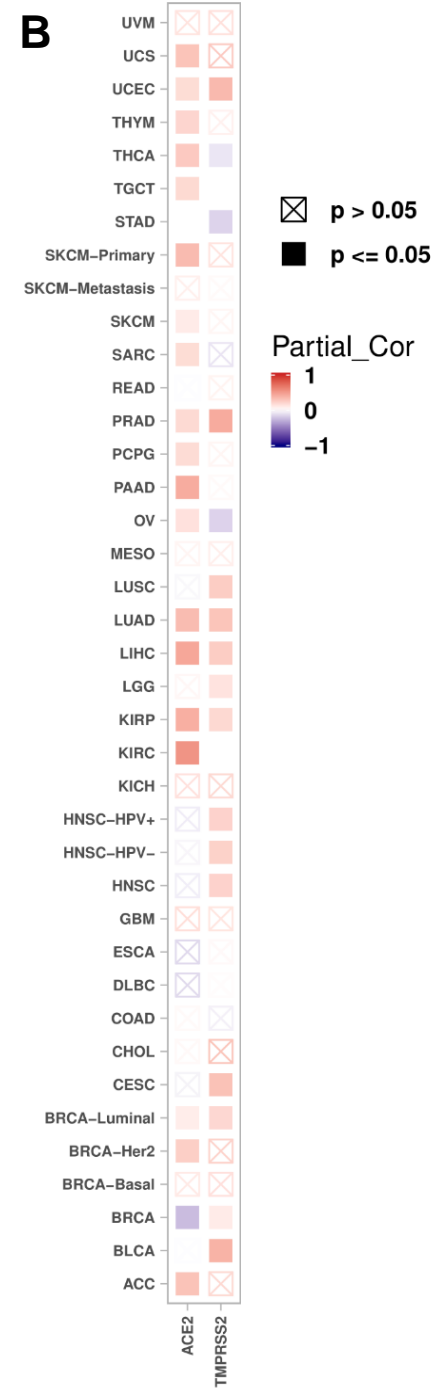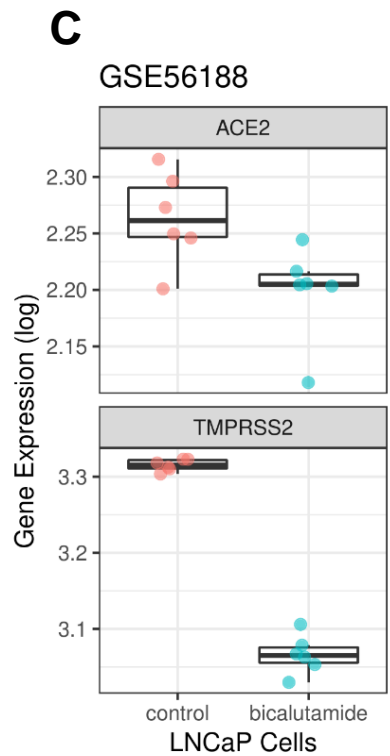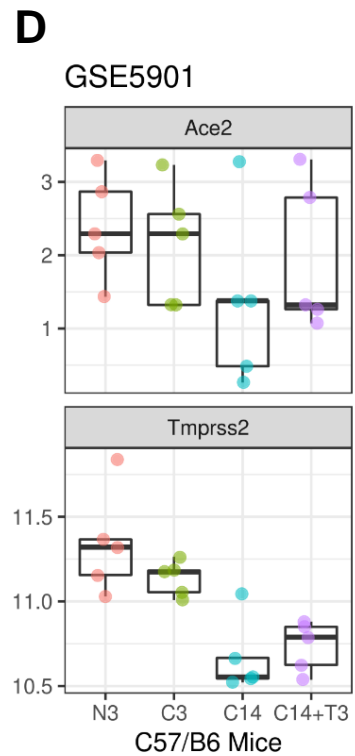

Table 1 Characteristics of all patients with COVID-19 included in the meta-analysis

| Author | Date | Deadline | Area | Average Age | Total Patients | Male | Female | Severe | smoking | Death | Comorbidities | Research Type | Quality | Reference |
|---|---|---|---|---|---|---|---|---|---|---|---|---|---|---|
| Qian G | 3/17 | 2/21 | China,Zhejiang | 50(5-96) | 91 | 37 | 54 | 9 | \ | 0 | 23 | Retrospective Research | 17 | 28 |
| Livingston E | 3/17 | 3/1 | Italy | 64 | 22512 | 13462 | 9050 | 6731 | \ | 1625 | \ | Report | 7 | 6 |
| Wang Y | 3/17 | 2/23 | China,Shenzhen | 49(2-69) | 55 | 22 | 23 | 2 | \ | 0 | 8 | Descriptive Research | 14 | 29 |
| KSID | 3/16 | 3/2 | Korea | 20-50 | 4212 | 1591 | 2621 | \ | \ | 22 | \ | Report | 15 | 13 |
| Su YJ | 3/14 | 2/11 | China,Taiwan | 56.6 | 10 | 7 | 3 | \ | \ | \ | \ | Retrospective Research | 11 | 30 |
| Dong X | 3/13 | \ | China,Tianjin | 48.62(16.83) | 135 | 72 | 63 | \ | \ | 3 | \ | Descriptive Research | 17 | 31 |
| Mizumoto K | 3/12 | 2/20 | Japan | \ | 634 | 321 | 313 | \ | \ | \ | \ | Descriptive Research | 16 | 47 |
| Deng L | 3/11 | 2/13 | China,Zhuhai | \ | 33 | 17 | 16 | \ | \ | \ | 19 | Retrospective Research | 16 | 32 |
| Zhou F | 3/11 | 1/31 | China,Wuhan | 56(46-67) | 191 | 119 | 72 | 119 | 11 | 54 | 91 | Retrospective Research | 15 | 10 |
| Wu Y | 3/10 | 2/2 | China, Wuhan | \ | 297 | 147 | 150 | 83 | \ | 17 | \ | Retrospective Research | 16 | 33 |
| Gao Q | 3/10 | 2/14 | China, Wuhan | 48（35-58.5) | 213 | 108 | 105 | 33 | \ | 7 | 19 | Retrospective Research | 14 | 34 |
| Chen X | 3/6 | 2/14 | China, Hubei | 46.0 (34.0-59.0) | 291 | 145 | 146 | 50 | \ | 2 | 93 | Observational Research | 16 | 35 |
| Zhang G | 3/6 | 2/10 | China, Wuhan | 55.0(39.0-66.5) | 221 | 108 | 113 | 55 | \ | 12 | 78 | Retrospective Research | 17 | 36 |
| Wu W | 3/6 | 2/6 | China, Guangdong | 43.1(17.1) | 21 | 10 | 11 | 7 | \ | \ | \ | Retrospective Research | 17 | 37 |
| Nicholas E | 3/6 | 2/27 | UK | 42.5（0.5-76) | 68 | 32 | 36 | \ | \ | \ | \ | Descriptive Research | 12 | 38 |
| Cao M | 3/6 | 2/25 | China,Shanghai | 50.1（16.3) | 198 | 101 | 97 | 19 | 11 | 1 | 69 | Retrospective Research | 17 | 39 |
| Zhao W | 3/3 | \ | China, Hunan | 44.44（12.3) | 101 | 56 | 45 | 14 | \ | \ | 30 | Retrospective Research | 14 | 40 |
| Young B | 3/3 | 2/3 | Singapore | 47(31-73) | 18 | 9 | 9 | 7 | \ | \ | 5 | Descriptive Research | 12 | 41 |
| Xiao F | 3/3 | 2/14 | China,Zhuhai | 43(0.83-78) | 73 | 41 | 32 | 4 | 9 | \ | 7 | Retrospective Research | 12 | 42 |
| Qi D | 3/3 | 2/16 | China, Chongqing | 48.0(20-80) | 267 | 149 | 118 | 50 | 53 | 4 | 41 | trospective and Descriptive Researc | 17 | 43 |
| Wang Y | 3/3 | 2/10 | China, Wuhan | \ | 110 | 48 | 62 | 38 | 26 | \ | 51 | Retrospective Research | 16 | 44 |
| Chen X | 3/3 | 2/27 | China, Wuhan | 64.6(18.1) | 48 | 37 | 11 | 27 | \ | 3 | \ | Retrospective Research | 15 | 45 |
| Cheng J | 3/2 | 1/2 | China, Henan | 46(36-59) | 1079 | 573 | 505 | 72 | \ | 11 | \ | Descriptive Research | 16 | 46 |
| Wu J | 2/29 | 2/14 | China,Jiangsu | 46.1(15.42) | 80 | 39 | 41 | 3 | \ | 0 | 38 | Descriptive Research | 16 | 48 |
| Li K | 2/29 | 2月 | China, Chongqing | 45.5(12.3) | 83 | 44 | 39 | 25 | \ | \ | 15 | Retrospective Research | 17 | 49 |
| Li J | 2/29 | 2/11 | China, Wuhan | 62.0（51.0-70.0) | 47 | 28 | 19 | 47 | \ | 1 | 30 | Retrospective Research | 16 | 50 |
| Guan W | 2/28 | 1/29 | China | 47(35-58) | 1099 | 637 | 459 | 173 | 261 | 15 | 261 | Retrospective Research | 15 | 8 |
| Tian S | 2/27 | 1/30 | China,Beijing | 47.5(1-94) | 262 | 127 | 135 | 46 | \ | 3 | \ | Retrospective Research | 17 | 51 |
| Liu Y | 2/27 | 2/1 | China, Wuhan | 55 (43-66) | 109 | 59 | 50 | 53 | \ | 31 | 76 | Retrospective Research | 17 | 52 |
| Xu Y | 2/25 | 2月 | China, Beijing | 43.9(16.8) | 50 | 29 | 21 | 13 | \ | \ | \ | Retrospective Research | 14 | 53 |
| Cao W | 2/25 | 2/20 | China,Xiangyang | \ | 128 | 60 | 68 | 21 | \ | \ | \ | Retrospective Research | 15 | 54 |
| Yang X | 2/24 | 1/26 | China, Wuhan | 59.7（13.3) | 52 | 35 | 17 | 52 | \ | 32 | 21 | Retrospective Research | 16 | 55 |
| Liu L | 2/23 | 2/3 | China, Chongqing | 45(34-51） | 51 | 32 | 19 | 7 | \ | 1 | 10 | Retrospective Research | 16 | 56 |
| Liu J | 2/22 | 1/24 | China, Wuhan | 48.7(13.9) | 40 | 15 | 25 | 13 | \ | \ | 14 | Retrospective Research | 16 | 57 |
| Zhang J | 2/19 | 3/2 | China,wuhan | 57（25-87) | 140 | 71 | 69 | 58 | 9 | \ | 90 | Retrospective Research | 15 | 58 |
| Xu X | 2/19 | 1/26 | China, Zhejiang | 41（32-52） | 62 | 36 | 26 | \ | \ | 0 | 20 | Retrospective Research | 15 | 59 |
| CDC | 2/17 | 2/11 | China | 30-69 | 44672 | 22981 | 21691 | 8255 | \ | 1023 | 5276 | Descriptive Research | 15 | 7 |
| Huang C | 2/15 | 1/2 | China, Wuhan | 49(41－58) | 41 | 30 | 11 | 13 | 3 | 6 | 13 | Retrospective Research | 16 | 60 |
| Wang D | 2/7 | 1/28 | China, Wuhan | 56(42-68) | 138 | 75 | 63 | 36 | \ | 6 | 64 | Retrospective Research | 17 | 5 |

**Table 2. Gender differences in the composition of patients with COVID-19**

| Author | Date | Area | Total Patients | Male | Female | Reference |
|---|---|---|---|---|---|---|
| Qian GQ | 3/17 | China,Zhejiang | 91 | 37 | 54 | 28 |
| Livingston E | 3/17 | Italy | 22512 | 13462 | 9050 | 6 |
| Wang Y | 3/17 | China,Shenzhen | 55 | 22 | 23 | 29 |
| KSID | 3/16 | Korea | 4212 | 1591 | 2621 | 13 |
| Su Y | 3/14 | China,Taiwan | 10 | 7 | 3 | 30 |
| Dong X | 3/13 | China,Tianjin | 135 | 72 | 63 | 31 |
| Deng L | 3/11 | China,Zhuhai | 33 | 17 | 16 | 32 |
| Zhou F | 3/11 | China,Wuhan | 191 | 119 | 72 | 10 |
| Wu Y | 3/10 | China, Wuhan | 297 | 147 | 150 | 33 |
| Gao Q | 3/10 | China, Wuhan | 213 | 108 | 105 | 34 |
| Chen X | 3/6 | China, Hubei | 291 | 145 | 146 | 35 |
| Zhang G | 3/6 | China, Wuhan | 221 | 108 | 113 | 36 |
| Wu W | 3/6 | China, Guangdong | 21 | 10 | 11 | 37 |
| Nicholas E | 3/6 | UK | 68 | 32 | 36 | 38 |
| Cao M | 3/6 | China,Shanghai | 198 | 101 | 97 | 39 |
| Zhao W | 3/3 | China, Hunan | 101 | 56 | 45 | 40 |
| Young B | 3/3 | Singapore | 18 | 9 | 9 | 41 |
| Xiao F | 3/3 | China,Zhuhai | 73 | 41 | 32 | 42 |
| Qi D | 3/3 | China, Chongqing | 267 | 149 | 118 | 43 |
| Wang Y | 3/3 | China, Wuhan | 110 | 48 | 62 | 44 |
| Chen X | 3/3 | China, Wuhan | 48 | 37 | 11 | 45 |
| Cheng J | 3/2 | China, Henan | 1079 | 573 | 505 | 46 |
| Mizumoto K | 3/1 | Japan | 634 | 321 | 313 | 47 |
| Wu J | 2/29 | China,Jiangsu | 80 | 39 | 41 | 48 |
| Li K | 2/29 | China, Chongqing | 83 | 44 | 39 | 49 |
| Li J | 2/29 | China, Wuhan | 47 | 28 | 19 | 50 |
| Guan W | 2/28 | China | 1099 | 637 | 459 | 8 |
| Tian S | 2/27 | China,Beijing | 262 | 127 | 135 | 51 |
| Liu Y | 2/27 | China, Wuhan | 109 | 59 | 50 | 52 |
| Xu Y | 2/25 | China, Beijing | 50 | 29 | 21 | 53 |
| Cao W | 2/25 | China,Xiangyang | 128 | 60 | 68 | 54 |
| Yang X | 2/24 | China, Wuhan | 52 | 35 | 17 | 55 |
| Liu L | 2/23 | China, Chongqing | 51 | 32 | 19 | 56 |
| Liu J | 2/22 | China, Wuhan | 40 | 15 | 25 | 57 |
| Zhang J | 2/19 | China,wuhan | 140 | 71 | 69 | 58 |
| Xu X | 2/19 | China, Zhejiang | 62 | 36 | 26 | 59 |
| CDC | 2/17 | China | 44672 | 22981 | 21691 | 7 |
| Huang C | 2/15 | China, Wuhan | 41 | 30 | 11 | 60 |
| Wang D | 2/7 | China, Wuhan | 138 | 75 | 63 | 5 |

**Table 3. Comparison of severe cases between male and female patients with COVID-19**

| Author | Date | Male | Male Severe | Female | Female Severe | P Value | Reference |
|---|---|---|---|---|---|---|---|
| Wu Y | 3/10 | 147 | 45 | 150 | 38 | \ | 33 |
| Gao Q | 3/10 | 108 | 22 | 105 | 11 | \ | 34 |
| Chen X | 3/6 | 145 | 27 | 146 | 23 | 0.629 | 35 |
| Zhang G | 3/6 | 108 | 35 | 113 | 20 | 0.011 | 36 |
| Wu W | 3/6 | 10 | 4 | 11 | 3 | 0.038 | 37 |
| Cao M | 3/6 | 101 | 17 | 97 | 2 | <0.001 | 39 |
| Qi D | 3/3 | 149 | 39 | 118 | 11 | <0.001 | 43 |
| Wang Y | 3/3 | 48 | 24 | 62 | 14 | 0.004 | 44 |
| Chen X | 3/3 | 37 | 24 | 11 | 3 | <0.001 | 45 |
| Zhao W | 3/1 | 56 | 8 | 45 | 6 | 0.89 | 40 |
| Li K | 2/29 | 44 | 15 | 39 | 10 | 0.402 | 49 |
| Guan W | 2/28 | 637 | 100 | 459 | 73 | \ | 8 |
| Tian S | 2/27 | 127 | 26 | 135 | 20 | 0.23 | 51 |
| Liu Y | 2/27 | 59 | 28 | 50 | 25 | 0.79 | 52 |
| Xu Y | 2/25 | 29 | 7 | 21 | 6 | \ | 53 |
| Cao W | 2/25 | 60 | 12 | 68 | 9 | >0.05 | 54 |
| Liu L | 2/23 | 32 | 3 | 19 | 4 | 0.109 | 56 |
| Liu J | 2/22 | 15 | 7 | 25 | 6 | 0.138 | 57 |
| Zhang J | 2/19 | 71 | 33 | 69 | 25 | 0.219 | 58 |
| Huang C | 2/15 | 30 | 11 | 11 | 2 | 0.24 | 60 |
| Wang D | 2/7 | 75 | 22 | 63 | 14 | 0.34 | 5 |

*Severe group in our study included ICU cases, severe case and critical cases.

**Table 4. Comparison of mortality between male and female patients with COVID-19**

| Author | Date | Male | Male Died | Female | Female Died | P Value | Reference |
|--------|------|------|-----------|--------|-------------|---------|-----------|
| KIDS   | 3/16 | 1591 | 13  | 2621  | 9   | \     | 13 |
| Zhou F | 3/11 | 119  | 38  | 72    | 16  | 0.15  | 10 |
| Wu Y   | 3/10 | 147  | 13  | 150   | 4   | \     | 33 |
| Gao Q  | 3/10 | 108  | 5   | 105   | 2   | \     | 34 |
| Zhang G| 3/6  | 108  | 7   | 113   | 2   | 0.681 | 36 |
| Cheng J| 3/2  | 573  | 7   | 505   | 4   | \     | 46 |
| Yang X | 2/24 | 35   | 21  | 17    | 11  | \     | 55 |
| CDC    | 2/17 | 22981| 653 | 21691 | 370 | \     | 7  |

# Supplementary Figure Legends

**Figure S1. Funnel plots for different outcomes.** Funnel plots for (A) male/female ratio, (2) odds ratio of severe cases, and (3) odds ratio of mortality.

**Figure S2. Bar charts representing cell-cell communications.** Intercellular communications between ACE2-expressing AT2 cells with other types of cells, with the ACE2 molecule functioning as (A) a receptor, and (B) a ligand.

**Figure S3. Single-cell transcriptomic profiling of normal prostate tissue samples.** (A) (B) scRNA-seq of normal prostate tissues from 3 healthy young donors (GSE117403). (A) Left, UMAP clustering of combined normal prostate tissue samples. Middle, feature plot showing ACE2 expression level of individual cells. Right, violin box plot showing the expression level of ACE2 among different cell populations. (B) Bar chart showing the percentage of ACE-positive cells in different epithelial cell clusters. (C) scRNA-seq of normal prostate tissues from 15 healthy donors aged 17-42 (GUDMAP 16-WPBW). Left, UMAP clustering of combined normal prostate tissue samples. Middle, feature plot showing ACE2 expression level of individual cells. Right, violin box plot showing the expression level of ACE2 among different cell populations.

**Figure S4. Single-cell transcriptomic profiling of testis tissue samples.** Clustering and ACE2 expression level of single testicular cells in (A) young adults, (B) spermatogonial stem cells, and (C) men during puberty. Left, clustering maps. Right, Feature maps showing ACE2 expression level of individual cells. SSCs, spermatogonial stem cells.

**Figure S5. Genome browser view of AR ChIP-seq and H3K27ac ChIP-seq tracks at the ACE2 locus in the C4-2B cell line.**

**Figure S6. Genome browser view of AR ChIP-seq, H3K27ac ChIP-seq, H3K4me1 ChIP-seq, and ATAC-seq tracks at the ACE2 locus in the LNCaP cell line.**

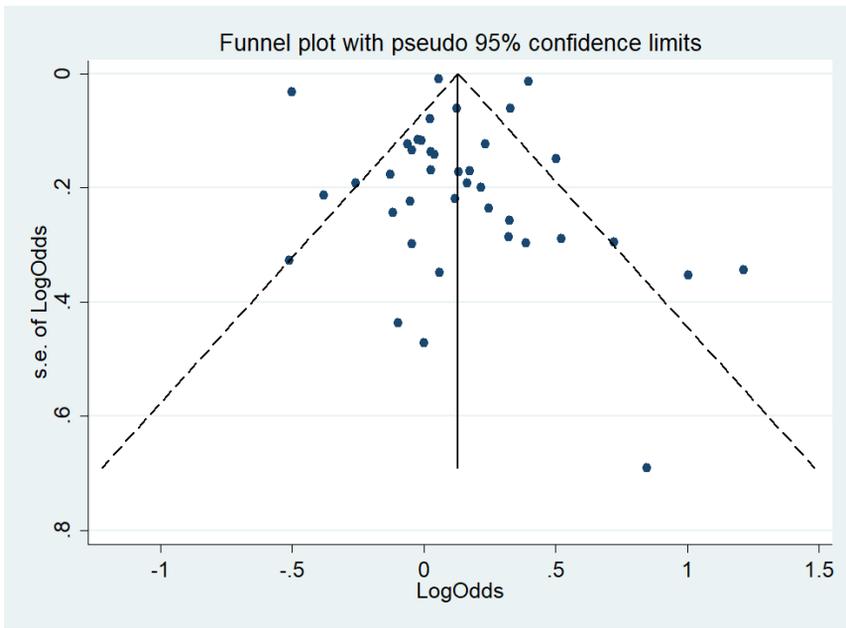
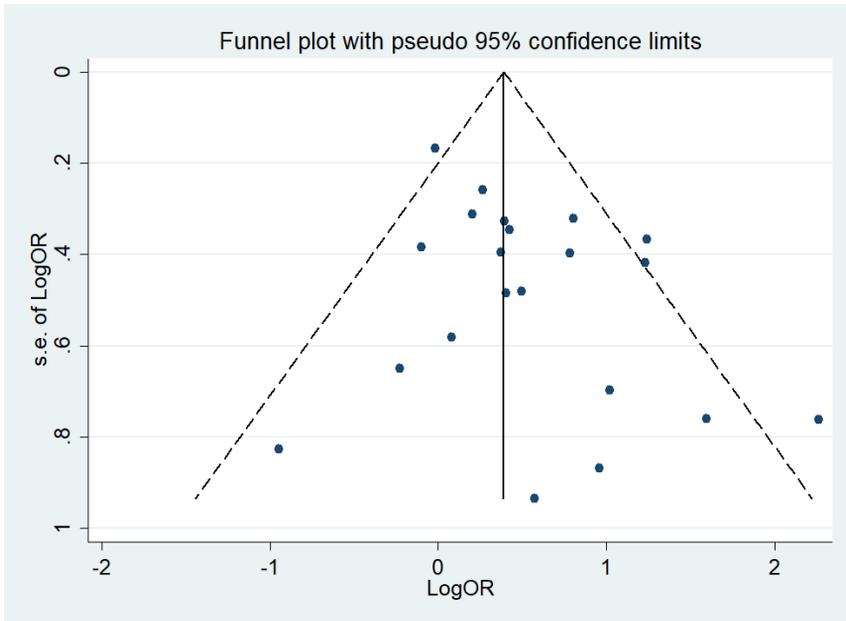
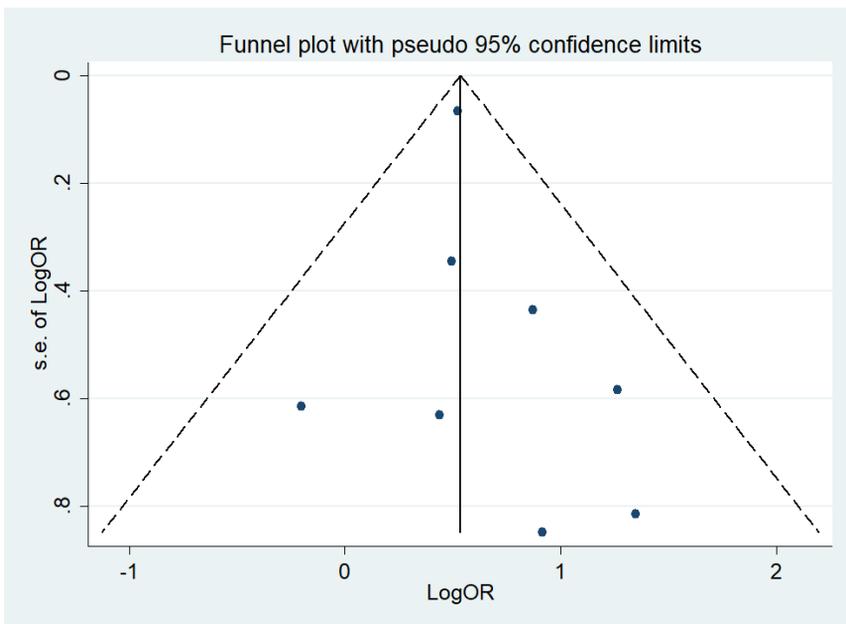

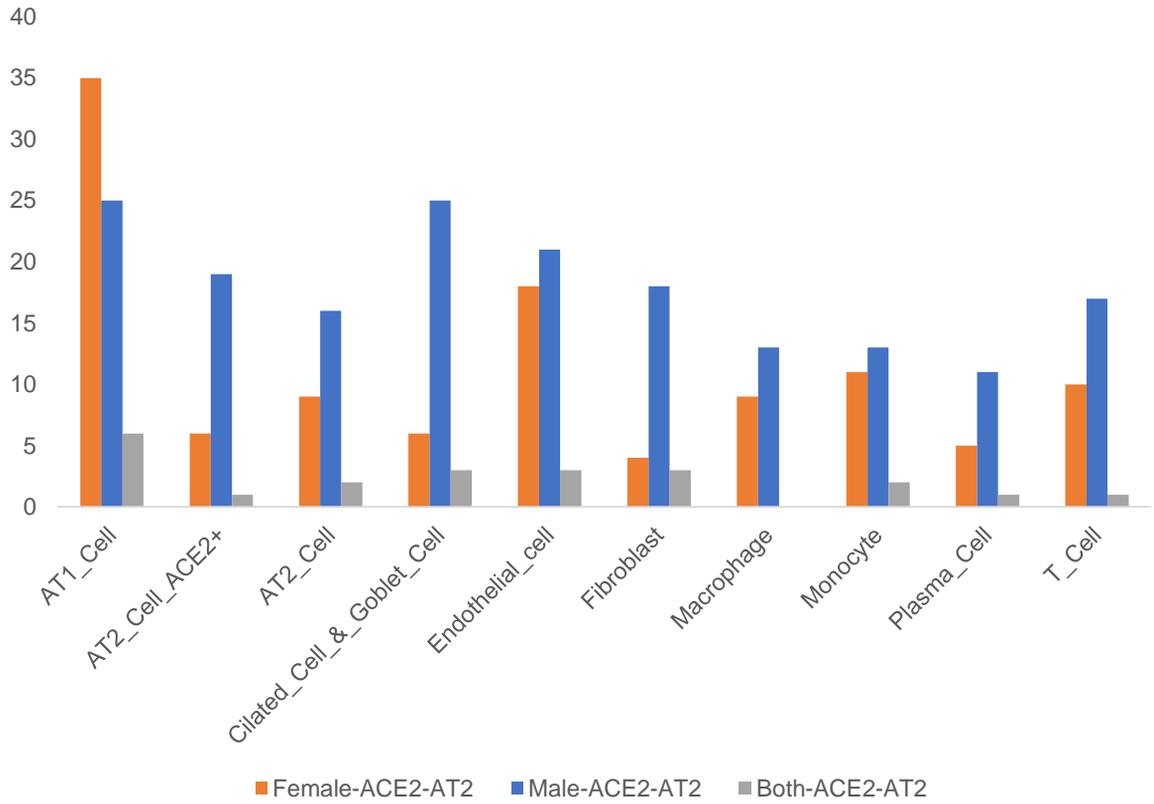

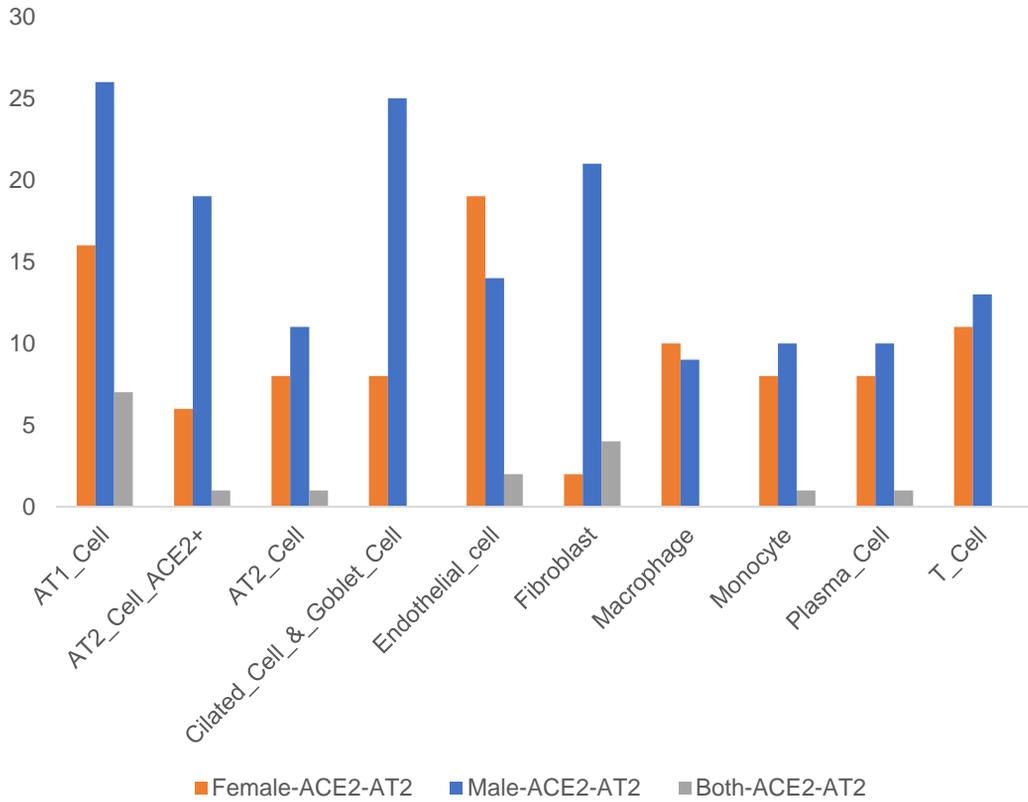

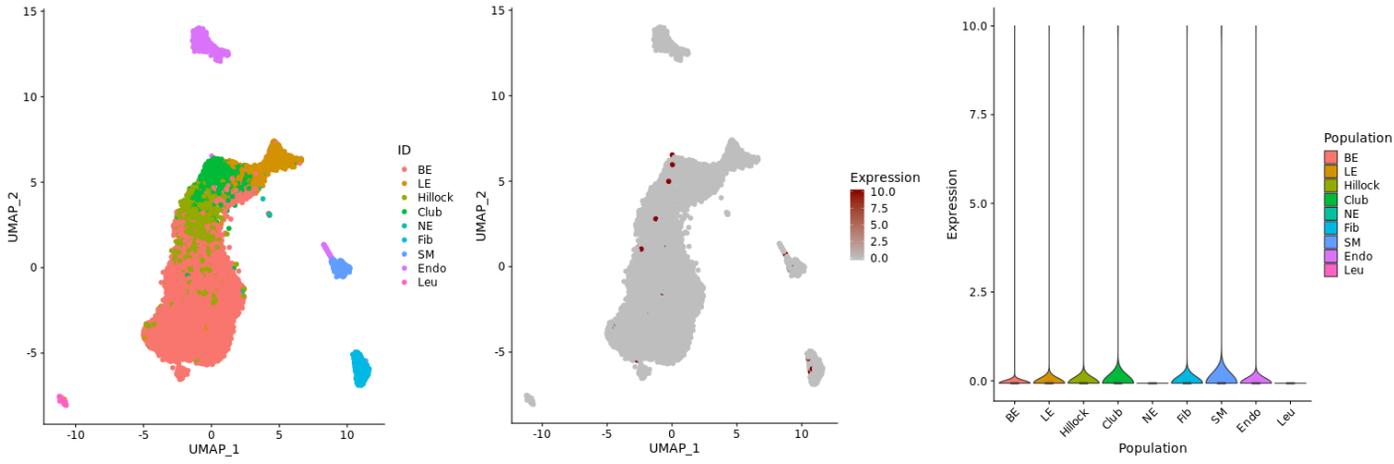

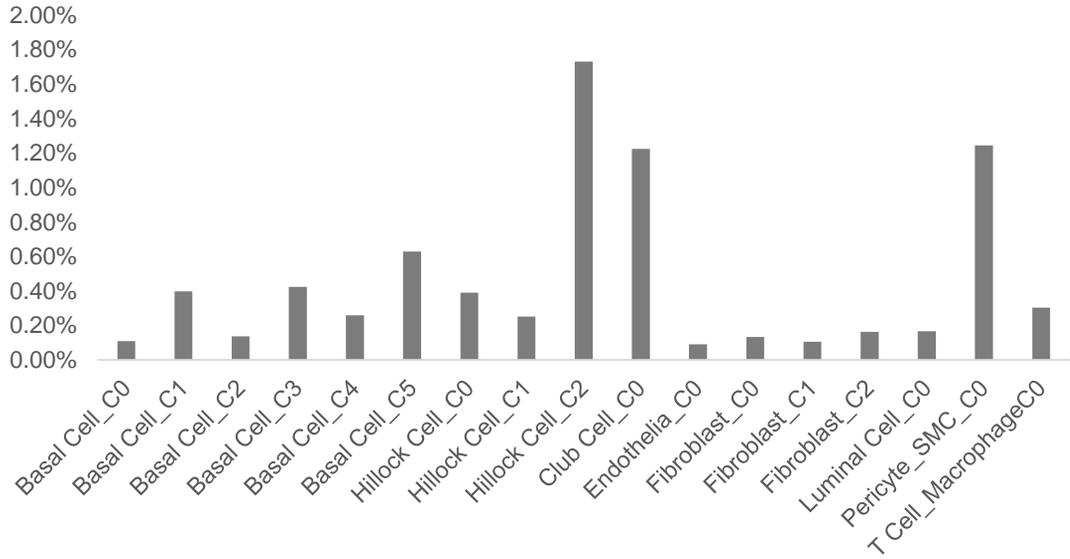

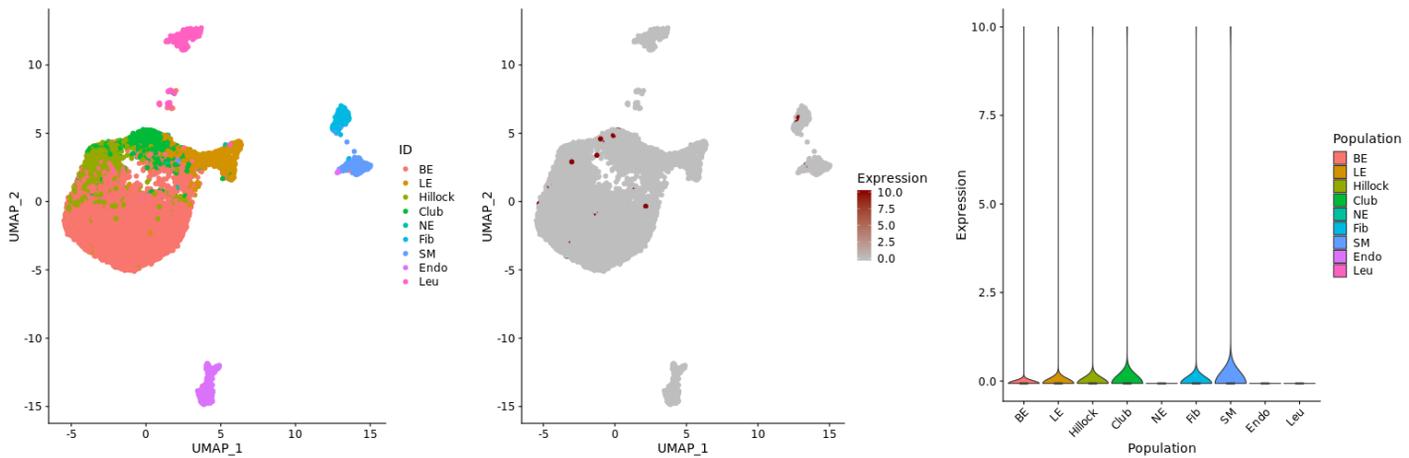

**A** Young Adult Atlas / ACE2

Clusters: D4, C3, E5, F6, B2, A1, H8, G7, J9, K11A, K11B, L12, J10

**B** Adult SSC States / ACE2

State 0, State 1, State 2, State 3, State 4

**C** Puberty Atlas / ACE2

C2_SYCT_3, C4_Sertoli_2, C4_Sertoli_1, C2_SYCT_2, C4_Sertoli_3, C3_STD_3, C3_STD_1, C2_SYCT_1, C4_Sertoli_4, C1_SGA_2, C3_STD_2, C8_Endo_1, C8_Endo_2, C1_SGA_1, C5_LeyMyo_1, C7_Macrophage, C5_LeyMyo_2, C5_LeyMyo_4, C6_Muscle

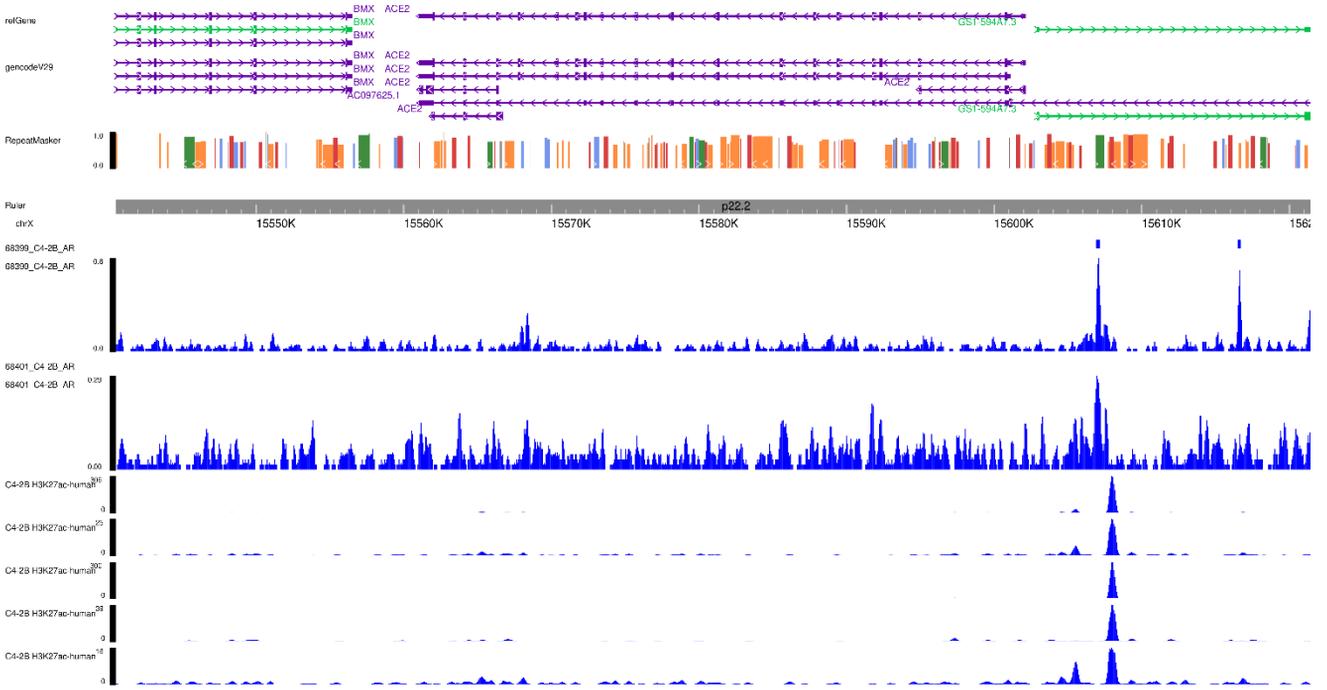

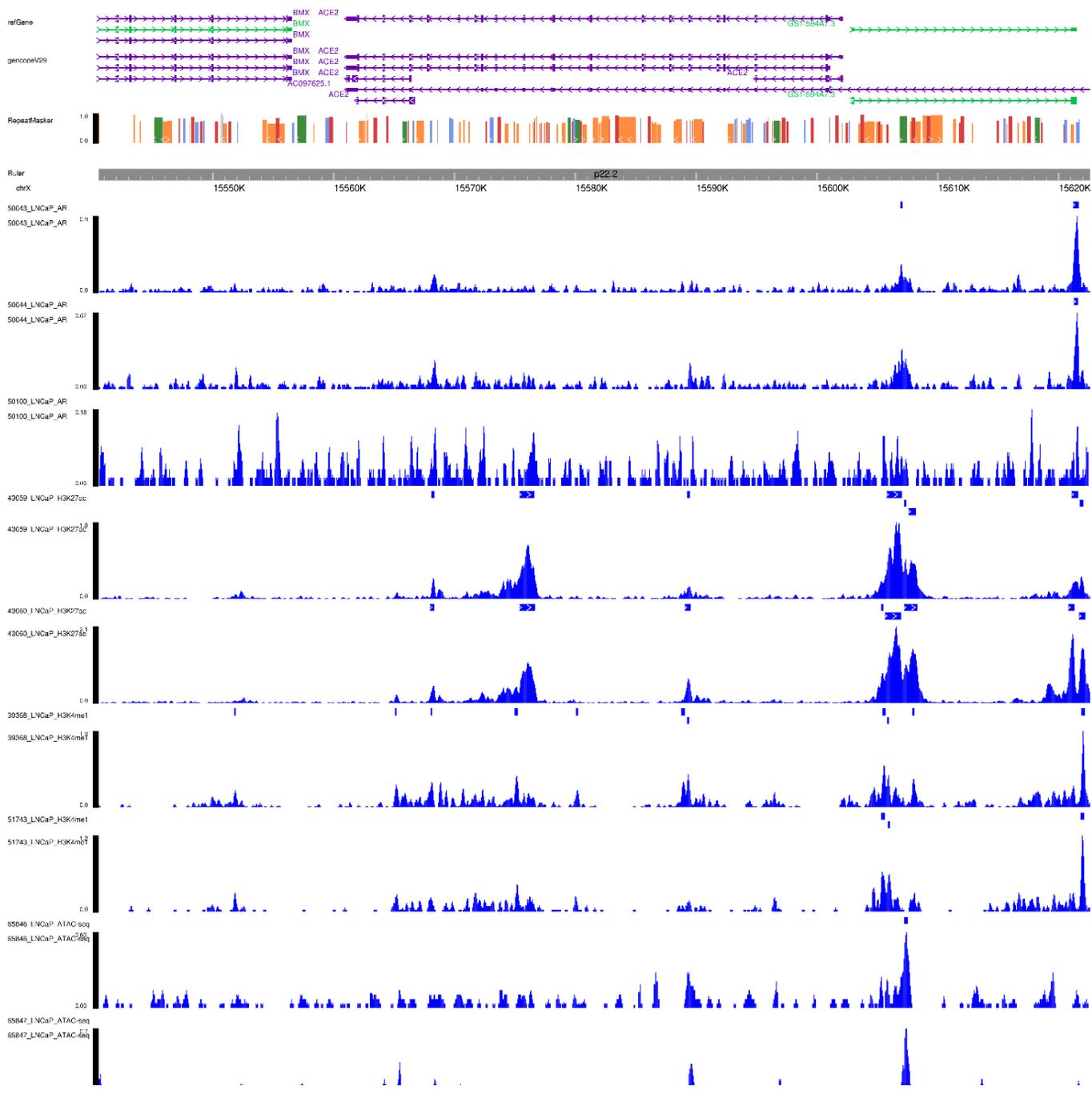

Table S1. Risk of bias and methodological quality of the included studies

| | | Introduction | Methods | | | | | | | | | | Results | | | | | Discussion | | Other | | |
|---|---|---|---|---|---|---|---|---|---|---|---|---|---|---|---|---|---|---|---|---|---|---|
| | | Q1 | Q2 | Q3 | Q4 | Q5 | Q6 | Q7 | Q8 | Q9 | Q10 | Q11 | Q12 | Q13 | Q14 | Q15 | Q16 | Q17 | Q18 | Q19 | Q20 | |
| Author | reference | 1.Were the aim | 2.Was the | 3.Was the | 4.Was the | 5.Was the | 6.Was the | 7.Were me | 8.Were the | 9.Were the | 10.Is it clea | 11.Were th | 12.Were th | 13.Does th | 14.If appro | 15.Were th | 16.Were th | 17.Discuss | 18.Were th | 19.Were th | 20.Was eth | Toal score |
| Qian G | 28 | Yes | Yes | Yes | Yes | Yes | Yes | ? | Yes | Yes | No | Yes | Yes | Yes | Yes | Yes | Yes | Yes | Yes | No | Yes | 17 |
| Livingston E | 6 | Yes | Yes | Yes | Yes | ? | ? | ? | ? | ? | No | ? | No | No | No | Yes | No | ? | Yes | ? | ? | 7 |
| Wang Y | 29 | Yes | Yes | Yes | Yes | Yes | Yes | ? | Yes | Yes | No | Yes | Yes | No | Yes | ? | Yes | ? | Yes | No | Yes | 14 |
| KSID | 13 | Yes | Yes | Yes | Yes | Yes | Yes | ? | Yes | Yes | No | Yes | Yes | Yes | No | ? | Yes | Yes | Yes | ? | Yes | 15 |
| Su YJ | 30 | Yes | Yes | Yes | Yes | Yes | Yes | ? | No | No | No | Yes | Yes | No | Yes | Yes | No | ? | ? | No | Yes | 11 |
| Dong X | 31 | Yes | Yes | Yes | Yes | Yes | Yes | ? | Yes | Yes | Yes | Yes | Yes | No | Yes | Yes | Yes | Yes | Yes | No | Yes | 17 |
| Mizumoto K | 47 | Yes | Yes | Yes | Yes | Yes | Yes | ? | Yes | Yes | No | Yes | Yes | No | Yes | Yes | Yes | Yes | Yes | No | Yes | 16 |
| Deng L | 32 | Yes | Yes | Yes | Yes | Yes | Yes | ? | ? | Yes | Yes | Yes | Yes | No | Yes | Yes | Yes | Yes | Yes | No | Yes | 16 |
| Zhou F | 10 | Yes | Yes | Yes | Yes | Yes | Yes | ? | Yes | Yes | Yes | Yes | Yes | No | ? | Yes | Yes | ? | Yes | No | Yes | 15 |
| Wu Y | 33 | Yes | Yes | Yes | Yes | Yes | Yes | ? | Yes | Yes | Yes | Yes | Yes | No | No | Yes | Yes | Yes | Yes | No | Yes | 16 |
| Gao Q | 34 | Yes | Yes | Yes | Yes | Yes | Yes | ? | Yes | No | No | Yes | Yes | No | Yes | Yes | ? | Yes | Yes | No | Yes | 14 |
| Chen X | 35 | Yes | Yes | Yes | Yes | Yes | Yes | ? | No | Yes | Yes | Yes | Yes | No | Yes | Yes | Yes | Yes | Yes | No | Yes | 16 |
| Zhang G | 36 | Yes | Yes | Yes | Yes | Yes | Yes | ? | Yes | Yes | Yes | Yes | Yes | No | Yes | Yes | Yes | Yes | Yes | No | Yes | 17 |
| Wu W | 37 | Yes | Yes | Yes | Yes | Yes | Yes | ? | Yes | Yes | Yes | Yes | Yes | No | Yes | Yes | Yes | Yes | Yes | No | Yes | 17 |
| Nicholas E | 38 | Yes | Yes | Yes | ? | Yes | Yes | ? | No | No | No | Yes | Yes | No | Yes | ? | Yes | Yes | Yes | No | Yes | 12 |
| Cao M | 39 | Yes | Yes | Yes | Yes | Yes | Yes | ? | Yes | Yes | Yes | Yes | Yes | No | Yes | Yes | Yes | Yes | Yes | No | Yes | 17 |
| Zhao W | 40 | Yes | Yes | Yes | Yes | Yes | Yes | ? | ? | ? | Yes | Yes | Yes | No | Yes | Yes | ? | Yes | Yes | No | Yes | 14 |
| Young B | 41 | Yes | Yes | Yes | Yes | Yes | Yes | ? | ? | ? | NO | Yes | Yes | No | No | Yes | Yes | ? | Yes | No | Yes | 12 |
| Xiao F | 42 | Yes | Yes | Yes | Yes | Yes | Yes | ? | ? | ? | NO | Yes | Yes | No | No | Yes | Yes | ? | Yes | No | Yes | 12 |
| Qi D | 43 | Yes | Yes | Yes | Yes | Yes | Yes | ? | Yes | Yes | Yes | Yes | Yes | No | Yes | Yes | Yes | Yes | Yes | No | Yes | 17 |
| Wang Y | 44 | Yes | Yes | Yes | Yes | Yes | Yes | ? | Yes | ? | Yes | Yes | Yes | No | Yes | Yes | Yes | Yes | Yes | No | Yes | 16 |
| Chen X | 45 | Yes | Yes | Yes | Yes | Yes | Yes | ? | Yes | Yes | Yes | Yes | Yes | No | Yes | Yes | Yes | No | ? | No | Yes | 15 |
| Cheng J | 46 | Yes | Yes | No | Yes | Yes | Yes | ? | Yes | Yes | Yes | Yes | Yes | No | Yes | Yes | Yes | Yes | Yes | No | Yes | 16 |
| Wu J | 48 | Yes | Yes | Yes | Yes | Yes | Yes | ? | Yes | Yes | Yes | Yes | Yes | No | Yes | Yes | Yes | No | Yes | No | Yes | 16 |
| Li K | 49 | Yes | Yes | Yes | Yes | Yes | Yes | ? | Yes | Yes | Yes | Yes | Yes | No | Yes | Yes | Yes | Yes | Yes | No | Yes | 17 |
| Li J | 50 | Yes | Yes | Yes | Yes | Yes | Yes | ? | No | Yes | Yes | Yes | Yes | No | Yes | Yes | Yes | Yes | Yes | No | Yes | 16 |
| Guan W | 8 | Yes | Yes | No | Yes | Yes | Yes | ? | No | Yes | Yes | Yes | Yes | No | Yes | Yes | Yes | Yes | Yes | No | Yes | 15 |
| Tian S | 51 | Yes | Yes | Yes | Yes | Yes | Yes | ? | Yes | Yes | Yes | Yes | Yes | No | Yes | Yes | Yes | Yes | Yes | No | Yes | 17 |
| Liu Y | 52 | Yes | Yes | Yes | Yes | Yes | Yes | ? | Yes | Yes | Yes | Yes | Yes | No | Yes | Yes | Yes | Yes | Yes | No | Yes | 17 |
| Xu Y | 53 | Yes | Yes | Yes | Yes | ? | Yes | ? | Yes | Yes | No | Yes | Yes | No | Yes | Yes | Yes | Yes | Yes | No | ? | 14 |
| Cao W | 54 | Yes | Yes | Yes | Yes | Yes | Yes | ? | Yes | Yes | Yes | Yes | Yes | No | ? | Yes | Yes | Yes | ? | No | Yes | 15 |
| Yang X | 55 | Yes | Yes | Yes | Yes | Yes | Yes | ? | Yes | Yes | Yes | Yes | Yes | No | ? | Yes | Yes | Yes | Yes | No | Yes | 16 |
| Liu L | 56 | Yes | Yes | Yes | Yes | ? | Yes | ? | Yes | Yes | Yes | Yes | Yes | No | Yes | Yes | Yes | Yes | Yes | No | Yes | 16 |
| Liu J | 57 | Yes | Yes | Yes | Yes | ? | Yes | ? | Yes | Yes | Yes | Yes | Yes | No | Yes | Yes | Yes | Yes | Yes | No | Yes | 16 |
| Zhang J | 58 | Yes | Yes | Yes | Yes | Yes | Yes | ? | Yes | Yes | Yes | Yes | Yes | No | ? | Yes | Yes | Yes | No | No | Yes | 15 |
| Xu X | 59 | Yes | Yes | Yes | Yes | ? | Yes | ? | Yes | Yes | Yes | Yes | Yes | No | ? | Yes | Yes | Yes | Yes | No | Yes | 15 |
| CDC | 7 | Yes | Yes | Yes | Yes | ? | Yes | ? | Yes | ? | Yes | Yes | Yes | No | Yes | Yes | Yes | Yes | Yes | No | Yes | 15 |
| Huang C | 60 | Yes | Yes | Yes | Yes | Yes | Yes | ? | ? | Yes | Yes | Yes | Yes | No | Yes | Yes | Yes | Yes | Yes | No | Yes | 16 |
| Wang D | 5 | Yes | Yes | Yes | Yes | Yes | Yes | ? | Yes | Yes | Yes | Yes | Yes | No | Yes | Yes | Yes | Yes | Yes | No | Yes | 17 |

## Table S2. Egger's regression test of each analysis.

|  | Coef. | Std. Err. | t | P> |t| | 95% CI | |
|---|---|---|---|---|---|---|
| Morbidity | -0.2573771 | 0.9040747 | -0.28 | 0.777 | -2.089206 | 1.574452 |
| Severity | 1.259753 | 0.6167905 | 2.04 | 0.055 | -0.0312042 | 2.550711 |
| Mortality | 0.358895 | 0.3756831 | 0.96 | 0.376 | -0.5603686 | 1.278159 |

## Table S3. Sensitivity analysis of the severity for 21 included studies

| Author | Date | Estimate OR | 95% CI | |
|---|---|---|---|---|
| Wu Y | 3/10 | 1.6678668 | 1.291166 | 2.1544712 |
| Gao Q | 3/10 | 1.6009253 | 1.252503 | 2.0462716 |
| Chen X | 3/6 | 1.6682906 | 1.29682 | 2.1461678 |
| Zhang G | 3/6 | 1.5895121 | 1.243146 | 2.0323834 |
| Wu W | 3/6 | 1.6275051 | 1.277919 | 2.0727232 |
| Cao M | 3/6 | 1.5475454 | 1.245357 | 1.9230609 |
| Qi D | 3/3 | 1.5309649 | 1.220754 | 1.9200046 |
| Wang Y | 3/3 | 1.5507388 | 1.23053 | 1.9542725 |
| Chen X | 3/3 | 1.5827113 | 1.253737 | 1.9980072 |
| Zhao W | 3/1 | 1.6524721 | 1.294761 | 2.10901 |
| Li K | 2/29 | 1.637333 | 1.279081 | 2.095926 |
| Guan W | 2/28 | 1.7173474 | 1.362389 | 2.1647866 |
| Tian S | 2/27 | 1.6450169 | 1.277818 | 2.1177359 |
| Liu Y | 2/27 | 1.6843767 | 1.32092 | 2.1478407 |
| Xu Y | 2/25 | 1.6603048 | 1.305324 | 2.1118217 |
| Cao W | 2/25 | 1.6307118 | 1.274032 | 2.0872478 |
| Liu L | 2/23 | 1.6617414 | 1.317398 | 2.0960901 |
| Liu J | 2/22 | 1.6055609 | 1.262431 | 2.0419543 |
| Zhang J | 2/19 | 1.6403354 | 1.275406 | 2.1096814 |
| Huang C | 2/15 | 1.6150899 | 1.269645 | 2.0545242 |
| Wang D | 2/7 | 1.6432891 | 1.280324 | 2.1091533 |
| Combined |  | 1.6256158 | 1.28406 | 2.0580247 |

## Table S4. Sensitivity analysis of the mortality for 8 included studies

| Author | Date | Estimate OR | 95% CI | |
|---|---|---|---|---|
| KIDS | 3/16 | 1.6962082 | 1.498673 | 1.9197793 |
| Zhou F | 3/11 | 1.7105864 | 1.510205 | 1.9375559 |
| Wu Y | 3/10 | 1.6939224 | 1.49752 | 1.916083 |
| Gao Q | 3/10 | 1.704739 | 1.507646 | 1.927598 |
| Zhang G | 3/6 | 1.700071 | 1.503474 | 1.9223759 |
| Cheng J | 3/2 | 1.7099531 | 1.511843 | 1.9340236 |
| Yang X | 2/24 | 1.7215263 | 1.522026 | 1.9471766 |
| CDC | 2/17 | 1.939573 | 1.307478 | 2.8772526 |
| Combined |  | 1.7082885 | 1.51129 | 1.9309661 |